\documentstyle[aps,preprint,tighten,floats,epsfig,amssymb,rotating]{revtex}
\begin{document}

\newcommand{\ebox}[2]{\epsfxsize=#1 \epsfbox[10 30 560 590]{#2}}
\newcommand{\be}{\begin{equation}}
\newcommand{\ee}{\end{equation}}
\newcommand{\rf}[4]{{\em {#1}} {\bf #2}, #3 (#4)}
\newcommand{\PR}[3]{\rf{Phys.\ Rev.}{#1}{#2}{#3}}
\newcommand{\pr}{Phys.\ Rev.\ }
\newcommand{\slh}{\!\!\!\slash}


\draft
\preprint{\vbox{\hfill \rm ADP-01-44/T476 \\
\null \hfill FSU-CSIT-02-01
}}

\title{\bf Overlap Quark Propagator in Landau Gauge}

\author{Fr\'{e}d\'{e}ric D.R.\ Bonnet$^{1}$,
Patrick O.\ Bowman$^{2}$,
Derek B.\ Leinweber$^{1}$,
Anthony G.\ Williams$^{1}$,
J.\ B.\ Zhang$^{1}$.}
\address{ $^1$ CSSM Lattice Collaboration, \\
Special Research Center for the Subatomic Structure of
Matter (CSSM) and Department of Physics and Mathematical Physics,
University of Adelaide 5005, Australia}
\address{ $^2$ Department of Physics and School for Computational Science
and Information
Technology, Florida State University, Tallahasse FL 32306, USA}
\date{\today}
\maketitle

\begin{abstract}

The properties of the quark propagator in Landau gauge in quenched
QCD are examined for the overlap quark action.  The overlap quark action
satisfies the Ginsparg-Wilson relation and as such provides an
exact lattice realization of chiral symmetry.  This in turn implies
that the quark action is free of ${\cal O}(a)$ errors.  We present results
using the standard Wilson fermion kernel in the overlap formalism on a 
$12^3\times 24$ lattice at a spacing of 0.125~fm.  We obtain the
nonperturbative momentum-dependent wavefunction renormalization function
$Z(p)$ and the nonperturbative mass function $M(p)$ for a variety
of bare masses.  We perform a simple extrapolation to the chiral limit
for these functions.  We clearly observe the dynamically generated
infrared mass and confirm the qualitative behavior found for the Landau
gauge quark propagator in earlier studies.  We attempt to extract the
quark condensate from the asymptotic behavior of the mass function in the
chiral limit. 

\end{abstract}

\section{INTRODUCTION}

Hadron correlators on the lattice provide a direct means of
calculating the physically observable properties of quantum chromodynamics
(QCD).  They are by construction color-singlet (i.e., gauge-invariant)
quantities.
Any finite, Boltzmann-distributed ensemble of gauge configurations has
negligible probability of containing two gauge-equivalent configurations.
Hence, there is a negligible probability of any gauge orbit
being represented more than once in the Monte-Carlo estimate of the
color-singlet hadron-correlator and this is the reason that there is
no need to gauge fix in such calculations.

On the other hand, calculations of high-energy
processes are carried out analytically with perturbative QCD,
where it is necessary to select a gauge.  Quark
models and QCD-inspired Dyson-Schwinger equation models\cite{DSE_Review}
are necessarily formulated in a particular gauge.  The usual
Fadeev-Popov gauge-fixing procedure is adequate for perturbative
QCD.  However in the nonperturbative infrared region standard gauge
choices, such as Landau gauge, have Gribov copies, i.e., there are
multiple gauge configurations on a given gauge orbit which satisfy
the gauge-fixing condition.  Since no finite ensemble will ever contain
two configurations from the same gauge orbit, Landau gauge on the lattice
actually corresponds to a gauge where there is a more or less random choice
between the Landau gauge Gribov copies on the represented gauge orbits in
the ensemble.  Before gauge fixing, the ensemble contains configurations
randomly located on their gauge orbits.  After Landau gauge fixing on the
lattice each configuration in the ensemble will typically settle on one
of the nearby Landau gauge Gribov copies.  This is the standard lattice
implementation of Landau gauge and the one that we consider in this work.  

In order to study the transition from non-perturbative to the perturbative
regime on the lattice we can study the gluon\cite{gluon_refs} and
quark\cite{becirevic,jon1,jon2,Bowman:2001xh} propagators and vertices such as
the quark-gluon vertex\cite{Skullerud:2001bu}. By studying the momentum
dependent quark mass function in the infrared region we can gain some
insights into the mechanism of dynamical chiral symmetry breaking
and the associated dynamical generation of mass.
Studying the ultraviolet behaviour of propagators at large momentum
is made difficult because of lattice artifacts causing the propagator 
to deviate strongly from its correct continuum behaviour in this
regime.  The method of tree-level correction was developed
and used successfully in gluon propagator studies\cite{gluon_refs}
and has recently been extended to the case of the quark
propagator\cite{jon1,jon2,Bowman:2001xh}.  Some related studies have
been performed for the case of domain-wall fermions\cite{Blum:2001sr}.
Detailed discusions of nonpertubative renormalization for lattice
operators can be found, e.g., in
Refs.~\cite{Martinelli:1994ty,Ciuchini:1995cd}

We present here results for the quark propagator obtained from
the overlap quark action and using an improved gauge action
and improved Landau gauge fixing.  The overlap action
is an exact realization of chiral symmetry on the lattice
and is necessarily ${\cal O}(a)$-improved.
In Secs.~\ref{action} and \ref{lattice}
we briefly introduce the improved gauge action and the lattice
quark propagator respectively.  In Sec.~\ref{overlapfermion} we
introduce the overlap quark propagator and
describe how it is calculated.
Our numerical results are presented in Sec.~\ref{numerical}
and finally in Sec.~\ref{conclusion} we give our summary and conclusions.

\section{IMPROVED GAUGE ACTION}
\label{action}

The tree-level ${\cal O}(a^2)$-improved action is defined as,
\begin{equation}
S_G=\frac{5\beta}{3}\sum_{\scriptstyle x\, \mu\, \nu \atop
\scriptstyle \nu > \mu} 
{\rm Re\, tr}(1 - P_{\mu \nu}(x))
-\frac{\beta}{12\, u_{0}^2}\sum_{\scriptstyle x\, \mu\, \nu\, \atop
\scriptstyle \nu > \mu}
{\rm Re\, tr}(1 - R_{\mu \nu}(x)),
\label{gaugeaction}
\end{equation}
where $P_{\mu \nu}$ and $R_{\mu \nu}$ are defined as
\begin{eqnarray}
\label{staplessq}
P_{\mu \nu}(x) & = & U_\mu(x)\, U_{\nu}(x+{\hat\mu})\, U^{\dagger}_{\mu}(x+{\hat\nu})
U^\dagger_{\nu}(x), \\
\label{staplesrect}
R_{\mu \nu}(x) & = & U_\mu(x)\, U_{\nu}(x+{\hat\mu})\, U_{\nu}(x+{\hat\nu}+{\hat\mu})\,
U^{\dagger}_{\mu}(x+2{\hat\nu})\, U^{\dagger}_{\nu}(x+{\hat\nu})\, U^\dagger_{\nu}(x) \nonumber \\
& + & U_\mu(x)\, U_{\mu}(x+{\hat\mu})\, U_{\nu}(x+2{\hat\mu})\,
U^{\dagger}_{\mu}(x+{\hat\mu}+{\hat\nu})\, U^{\dagger}_{\mu}(x+{\hat\nu})\, U^\dagger_{\nu}(x).
\end{eqnarray}
The link product $R_{\mu \nu}(x)$ denotes the rectangular $1\times2$
and $2\times1$ plaquettes. The mean link, $u_0$, is the tadpole (or
mean-field) improvement
factor that largely corrects for the quantum renormalization of the
coefficient for the rectangles relative to the plaquette.
We employ the plaquette measure for the mean link
\begin{equation}
u_0=\left(\frac{1}{3}{\rm Re\, tr}\left< P_{\mu \nu}(x) \right>
\right)^{1/4} \, ,
\label{uzero}
\end{equation}
where the angular brackets indicate averaging over $x$, $\mu$, and
$\nu$. 

Gauge configurations are generated using the
Cabbibo-Marinari~\cite{Cab82} pseudo--heat--bath algorithm with three
diagonal $SU_{c}(2)$ subgroups cycled twice. Simulations are performed 
using a parallel algorithm on a Sun Cluster composed of 40 nodes and
on a Thinking Machines Corporations (TMC) CM-5 both with appropriate
link partitioning.  We partition the
link variables according to the algorithm described in Ref.~\cite{mask}.
We use 50 configurations generated on a $12^3\times{24}$ lattice at
$\beta=4.60$, selected after 5000 thermalization sweeps from a
cold start and every 500 sweeps thereafter with a fixed mean--link value.
Lattice parameters are summarized in Table~\ref{simultab}.  The lattice
spacing is determined from the static quark potential with a string
tension $\sqrt{\sigma}=440$~MeV\cite{zanotti}.

\begin{table}[b]
\caption{Lattice parameters.}
\begin{tabular}{cccccccc}
Action &Volume &$N_{\rm{Therm}}$ & $N_{\rm{Samp}}$ &$\beta$ &$a$ (fm) & $u_{0}$ & Physical Volume (fm$^4$)\\
\hline
Improved       & $12^3\times{24}$ & 5000 & 500 & 4.60 & 0.125  & 0.88888 & $1.5^3\times{3.00}$ \\
\end{tabular}
\label{simultab}
\end{table}

The gauge field configurations are gauge fixed to the Landau gauge using a
Conjugate Gradient Fourier Acceleration~\cite{cm} algorithm with an accuracy
of $\theta\equiv\sum\left|\partial_{\mu}A_{\mu}(x)\right|^{2}<10^{-12}$.
We use an improved gauge-fixing scheme to minimize
gauge-fixing discretization errors.  A discussion of the functional and
method for improved Landau gauge fixing can be found in Ref.~\cite{bowman}.

\section{Quark Propagator on the Lattice}
\label{lattice}

In a covariant gauge in the continuum the renormalized Euclidean
space quark propagator must have the form
\begin{eqnarray}
S(\zeta;p)=\frac{1}{i {p \slh} A(\zeta;p^2)+B(\zeta;p^2)}
=\frac{Z(\zeta;p^2)}{i{p\slh}+M(p^2)}\, ,
\label{ren_prop}
\end{eqnarray}
where $\zeta$ is the renormalization point and where the renormalization
point boundary conditions are
$Z(\zeta;\zeta^2)\equiv 1$ and $M(\zeta^2)\equiv m(\zeta)$ and where
$m(\zeta)$ is the renormalized quark mass at the renormalization point.  
Since the gauge fixing condition has no preferred direction in color
space, the quark propagator must be diagonal in color space, i.e.,
$S^{ij}(\zeta;p) = S(\zeta;p)\delta^{ij}$ where $\delta^{ij}$ is the
$3\times 3$, $SU(3)_c$ identity matrix.
The functions $A(\zeta;p^2)$ and $B(\zeta;p^2)$, or alternatively
$Z(\zeta;p^2)$ and $M(p^2)$,  contain all of the nonperturbative information
of the quark propagator.  Note that $M(p^2)$ is renormalization
point independent, i.e., since $S(\zeta;p)$ is multiplicatively
renormalizable all of the renormalization-point dependence is carried by 
$Z(\zeta;p^2)$.  For sufficiently large momenta the effects of dynamical
chiral symmetry breaking become negligible for nonzero current quark masses,
i.e., for large $\zeta$ and $m_\zeta\neq 0$ we have
$m(\zeta)\to m_\zeta$ where $m_\zeta$ is the usual
current quark mass of perturbative QCD at the renormalization point $\zeta$. 
When all interactions for the quarks are turned off, i.e., when the gluon
field vanishes, the quark propagator has its tree-level form
\begin{equation}
S^{(0)}(p)=\frac{1}{i{p\slh}+m^0} \, ,
\end{equation}
where $m^0$ is the bare quark mass.  When the interactions with the
gluon field are turned on we have
\begin{equation}
S^{(0)}(p) \to S^{\rm bare}(a;p) = Z_2(\zeta;a) S(\zeta;p) \, ,
\label{tree_bare_ren}
\end{equation}
where $a$ is the regularization parameter (i.e., the lattice spacing here)
and $Z_2(\zeta;a)$ is the quark wave-function renormalization constant
chosen so as to ensure $Z(\zeta;\zeta^2)=1$. For simplicity of notation we suppress the $a$-dependence
of the bare quantities. 

On the lattice we expect the bare quark propagators, in momentum space,
to have a similar form as in the continuum~\cite{becirevic,jon1,jon2}, except
that the $O(4)$ invariance is replaced by a 4-dimensional hypercubic
symmetry on an isotropic lattice.
Hence, the inverse lattice bare quark propagator takes the general form
\be
(S^{\rm bare})^{-1}(p)\equiv
{i\left(\sum_{\mu}C_{\mu}(p)\gamma_{\mu}\right)+B(p)}.
\label{invquargen}
\ee
We use periodic boundary conditions in the spatial directions
and anti-periodic in the time direction.
The discrete momentum values for a
lattice of size $N^{3}_{i}\times{N_{t}}$, with $n_i=1,..,N_i$ and $n_t=1,..,N_t$, are
\begin{eqnarray}
p_i=\frac{2\pi}{N_{i}a}\left(n_i-\frac{N_i}{2}\right),\hspace{1cm}{\rm{and}}\hspace{1cm}p_t=\frac{2\pi}{N_{t}a}\left(n_t-\frac{1}{2}-\frac{N_t}{2}\right).
\label{dismomt}
\end{eqnarray}
Defining the bare lattice quark propagator as
\begin{equation}
S^{\rm bare}(p)\equiv
{-i\left(\sum_{\mu}{\cal{C}}_{\mu}(p)\gamma_{\mu}\right)+{\cal{B}}(p)}\, ,
\end{equation}
we perform a spinor and color trace to identify
\begin{eqnarray}
{\cal{C}}_{\mu}(p)=\frac{i}{4N_c}{\rm tr}[\gamma_\mu{S^{\rm bare}(p)}]
\hspace{1.0cm}&{\rm{and}}&\hspace{1.0cm}
{\cal{B}}(p)=\frac{1}{4N_c}{\rm tr}[S^{\rm bare}(p)] \, .
\label{curlyCandB}
\end{eqnarray}
The inverse propagator is
\begin{eqnarray}
(S^{\rm bare})^{-1}(p)&=&\frac{1}{{-i\left(\sum_{\mu}{\cal{C}}_{\mu}(p)\gamma_{\mu}\right)+{\cal{B}}(p)}}\nonumber\\
&=&\frac{{i\left(\sum_{\mu}{\cal{C}}_{\mu}(p)\gamma_{\mu}\right)+{\cal{B}}(p)}}{{\cal{C}}^{2}(p)+{\cal{B}}^{2}(p)}\, ,
\end{eqnarray}
where ${\cal{C}}^{2}(p)=\sum_{\mu}({\cal{C}}_{\mu}(p))^{2}$. From Eq.~(\ref{invquargen}) we identify
%
\begin{equation}
C_{\mu}(p) = \frac{{\cal{C}}_{\mu}(p)}{{\cal{C}}^{2}(p)+{\cal{B}}^{2}(p)}
\hspace{1.0cm}{\rm{and}}\hspace{1.0cm}
B(p) = \frac{{\cal{B}}(p)}{{\cal{C}}^{2}(p)+{\cal{B}}^{2}(p)} \, .
\label{cmup_bp}
\end{equation}

\subsection{Tree-level correction}
\label{latticemz}

At tree--level, i.e., when all the gauge links are set to the identity, the
inverse bare lattice
quark propagator becomes the tree-level version of
Eq.~(\ref{invquargen})
\be
(S^{(0)})^{-1}(p)\equiv
{i\left(\sum_{\mu}C_{\mu}^{(0)}(p)\gamma_{\mu}\right)+B^{(0)}(p)}\,.
\label{treeinvpro}
\ee
We calculate $(S^{(0)})(p)$ directly by setting the links to unity in the
coordinate space quark propagator and taking its Fourier transform

It is then possible to identify the appropriate kinematic lattice
momentum directly from the definition
\begin{eqnarray}
q_\mu\equiv C_{\mu}^{(0)}(p)&=&\frac{{\cal{C}}_{\mu}^{(0)}(p)}{({\cal{C}}^{(0)}(p))^{2}+({\cal{B}}^{(0)}(p))^{2}} \, .
\label{latmomt}
\end{eqnarray}
This is the starting point for the general approach to tree-level
correction developed in earlier studies of the gluon
propagator\cite{gluon_refs} and the quark
propagator\cite{jon1,jon2,Bowman:2001xh}.

Having identified
the appropriate kinematical lattice momentum $q$, we can now
define the bare lattice propagator as
\begin{equation}
S^{\rm bare}(p)
\equiv \frac{1}{i{q\slh}A(p)+B(p)}
=\frac{Z(p)}{i{q\slh}+M(p)} = Z_2(\zeta;a) S(\zeta;p)
\end{equation}
and the lattice version of the renormalized propagator in
Eq.~(\ref{ren_prop})
\begin{equation}
S(\zeta;p)
\equiv \frac{1}{i{q\slh}A(\zeta;p)+B(\zeta;p)}
=\frac{Z(\zeta;p)}{i{q\slh}+M(p)} \, .
\end{equation}

The general approach to tree-level
correction\cite{gluon_refs,jon1,jon2,Bowman:2001xh}
utilizes the fact that QCD
is asymptotically free and so it is the difference of bare quantities 
from their tree-level form on the lattice that contains the best
estimate of the nonperturbative
information.  For example, multiplicative tree-level correction
for $Z(p)$ and $M(p)$ has the form
\begin{eqnarray}
Z^{(c)}(p)=\frac{Z(p)}{Z^{(0)}(p)}1
\hspace{1cm}{\rm{and}}\hspace{1cm}
M^{(c)}(p)=\frac{M(p)}{M^{(0)}(p)}m^0 \, .
\label{acbc}
\end{eqnarray}
The identification of the kinematical variable $q$ ensures that
$A^{\rm (0)}(p)=1/Z^{\rm (0)}(p)=1$ by construction and so 
$Z(p)=Z^{(c)}(p)$ and is already tree-level corrected.
For overlap quarks we will see that $M^{(0)}(p)=m^0$ and so the
mass function satisfies $M(p)=M^{(c)}(p)$ and needs no tree-level
correction either.  This feature is a major advantage of the overlap
formalism.

\section{OVERLAP FERMIONS.}
\label{overlapfermion}

The overlap fermion formalism~\cite{neuberger0,neuberger1,neuberger2,Narayanan:1994gw,Niedermayer:1998bi,edwards2,Luscher:1998pq,zhang,heller,Giusti}
realizes an exact chiral
symmetry on the lattice and is automatically ${\cal O}(a)$ improved, since
any ${\cal O}(a)$ error would necessarily violate chiral
symmetry\cite{Niedermayer:1998bi}. 
The massless coordinate-space overlap-Dirac operator can be written in
dimensionless lattice units as\cite{edwards2}
\begin{equation}
   D(0) = \frac{1}{2}\left[1 + \gamma_5 H_a\right] \, ,
\label{D0_definition}
\end{equation}
where $H_a$ is an Hermitian operator that depends on the background gauge
field and has eigenvalues $\pm 1$.
Any such $D(0)$ is easily seen to satisfy the Ginsparg--Wilson
relation~\cite{ginsparg}
\begin{eqnarray}
\{\gamma_{5},D(0)\}=2D(0)\gamma_{5}D(0)
\label{ginspargrel}
\end{eqnarray}
and, provided that its Fourier transform at low momenta is proportional
to the momentum-space covariant derivative, it will satisfy a deformed
lattice realization of chiral symmetry.
It immediately follows from Eq.~(\ref{D0_definition}) that
\begin{equation}
D^\dagger(0)D(0)=D(0)D^\dagger(0)=\frac{1}{2}[D^\dagger(0)+D(0)]
\end{equation}
and that
\begin{equation}
D^\dagger(0)=\gamma_5 D(0)\gamma_5 \, .
\end{equation}
It also follows easily that $\{\gamma_5,D^{-1}(0)\}=2\gamma_5$ and
by defining $\tilde D^{-1}(0)\equiv [D^{-1}(0)-1]$ we see
that the Ginsparg-Wilson relation can also be expressed in the
form
\begin{equation}
\{\gamma_5,\tilde D^{-1}(0)\}=0 \, .
\end{equation}

The standard choice of $H_a(x,y)$ is $H_a=\epsilon(H_w)\equiv 
H_w/|H_w| = H_w/(H_w^\dagger H_w)^{1/2}$, where
$H_w(x,y)=\gamma_5 D_w(x,y)$ is the Hermitian Wilson-Dirac
operator and where $D_w$ is the usual Wilson-Dirac operator 
on the lattice.  However, in the overlap formalism the Wilson mass
parameter $m_w$ is the negative of what it is for standard Wilson
fermions and at tree-level must satisfy $0 <m_wa <2$.
In the overlap formalism $m_w$ is an intermediate
lattice regularization parameter, it is not the bare quark mass.  
When interactions are present, we must have $m_1a<m_wa$ in order
that the Wilson operator has zero-crossings and, in turn, that
$D(0)$ has nontrivial topological charge.  Numerical studies have found
that $m_1\simeq m_c$, where $m_c$ is the usual critical mass for Wilson
fermions\cite{Edwards:1998sh}.  The constraint $m_wa<2$ at tree level
arises from the fact that Wilson doublers reappear above this point.
In summary, we use here $H_w(-m_w)=\gamma_5\,D_w(-m_w) $.

Recall that the standard Wilson-Dirac operator can be written as
\begin{eqnarray}
D_{w}(x,y)
&=& \left[ (-m_wa)+4r\right]\delta_{x,y}
-\frac{1}{2}\sum_\mu\left\{(r-\gamma_\mu)
U_\mu(x)\delta_{y,x+\hat\mu}+(r+\gamma_\mu)
U^\dagger_\mu(x-a\hat{\mu})\delta_{y,x-\hat\mu}\right\} \nonumber \\
&=& \frac{1}{2\kappa^{\rm st}} \left[\delta_{x,y}
-\kappa^{\rm st}\sum_\mu
\left\{(r-\gamma_\mu)
U_\mu(x)\delta_{y,x+\hat\mu}+(r+\gamma_\mu)
U^\dagger_\mu(x-a\hat{\mu})\delta_{y,x-\hat\mu}
\right\}
\right] \, ,
\label{standard_diracop}
\end{eqnarray}
where the negative Wilson mass term $(-m_wa)$ is then defined by
$(-m_wa)+4r=1/2\kappa^{\rm st}$ or equivalently
\begin{equation}
\kappa^{\rm st}\equiv\frac{1}{2(-m_wa)+(1/\kappa_c)}
\label{standard_kappa_defn}
\end{equation}
and where $\kappa_c$ throughout this work is the tree-level critical
$\kappa$, i.e., $\kappa_c=1/(8r)$.

In the present work we use
the mean-field improved Wilson-Dirac operator, which can be written as
\begin{equation}
D_{w}(x,y)
= \frac{u_0}{2\kappa} \left[\delta_{x,y}
-\kappa\sum_\mu
\left\{(r-\gamma_\mu)
\frac{U_\mu(x)}{u_0}\delta_{y,x+\hat\mu}+(r+\gamma_\mu)
\frac{U^\dagger_\mu(x-a\hat{\mu})}{u_0}\delta_{y,x-\hat\mu}
\right\}
\right] \, .
\label{diracop}
\end{equation}
We see that this is equivalent to the standard Wilson-Dirac operator
with the identification of the mean-field improved coefficient
$\kappa\equiv \kappa^{\rm st} u_0$.  It is $U/u_0$ that has a more convergent
expansion around the identity than the links $U$ themselves.
The negative Wilson mass $(-m_wa)$ is then related to this improved $\kappa$ by
\begin{equation}
\kappa\equiv\frac{u_0}{2(-m_wa)+(1/\kappa_c)} \, .
\label{kappa_defn}
\end{equation}
The Wilson parameter is typically chosen to be $r=1$ and
we will also use $r=1$ here in our numerical
simulations.

For this mean-field improved Wilson-Dirac choice we then have
 
\begin{equation}
D(0) = \frac{1}{2}\left[1 + D_w\left(D_w^\dagger D_w\right)^{-1/2}
       \right] \, .
\label{D0_Wilson}
\end{equation}
In coordinate space the Wilson-Dirac operator has the form
$D_w=\nabla\!\!\!\!\slash + (r/2)\Delta + (-m_wa)$, where
$\nabla_\mu$ is the symmetric dimensionless lattice finite difference
operator, and $\Delta$ is the dimensionless lattice Laplacian operator.
Recall that the Wilson mass term is $(-m_wa)$ here.
Setting the links to the identity gives
\begin{equation}
D_w=(1/2)(\stackrel{\leftarrow}{\partial}_\mu+
\stackrel{\rightarrow}{\partial}_\mu )\gamma_\mu
+(r/2) (-\stackrel{\leftarrow}{\partial}_\mu
\stackrel{\rightarrow}{\partial}_\mu ) + (-m_wa)\, ,
\end{equation}
where the partial derivatives are the forward and backward lattice
finite difference operators.
Hence we have from Eq.~(\ref{D0_Wilson})
that
\begin{eqnarray}
D(0)&=& \frac{1}{2}\left[1 + 
  \frac{\nabla\!\!\!\!\slash+(r/2) \Delta  - m_wa}
       {\sqrt{(m_wa)^2 +{\cal O}(\partial^2)}}\right]
  \\ \nonumber
      &\to&  \frac{\nabla\!\!\!\!\slash}{2m_wa} \, ,
\end{eqnarray}
where the last line is a limit approached when operating on very smooth
functions such that only first powers of derivatives are kept.
The reason for needing a negative Wilson mass $(-m_wa)$ is now apparent,
i.e., it is needed to cancel the 1 in $D(0)$.
We see that at sufficiently fine lattice spacings and for $pa\ll 1$ 
\begin{equation}
D_c(0)\equiv (2m_w)D(0)\, ,
\end{equation}
where $D_c(0)$ in the continuum limit becomes the usual fermion covariant derivative contracted with
the $\gamma$-matrices, i.e., $D_c(0)\to D\!\!\!\!\slash$ as $a\to 0$.

The massless overlap quark propagator is given by
\begin{equation}
S^{\rm bare}(0) \equiv \tilde D_c^{-1}(0)
                \equiv D_c^{-1}(0)-\frac{1}{2m_w}
                = \frac{1}{2m_w}\left[ D^{-1}(0)-1\right]
                = \frac{1}{2m_w}\tilde D^{-1}(0) \, .
\label{massless_quark}
\end{equation}
This definition of the massless overlap quark propagator follows
from the overlap formalism\cite{Narayanan:1994gw}
and ensures that the massless quark propagator anticommutes with
$\gamma_5$, i.e., 
$\{\gamma_5,S^{\rm bare}(0)\}=0$ just as it does in the continuum
\cite{edwards2}.  At tree-level the momentum-space form of the massless
propagator defines the kinematic lattice momentum $q$, 
i.e., we set the links to one
such that we have for the momentum-space massless quark
propagator 
\begin{equation}
S^{\rm bare}(0,p)\equiv \tilde D_c^{-1}(0,p)
       \to S^{(0)}(0,p)=\frac{1}{iq\slh} \, , 
\end{equation}
where recall that $p$ is the discrete lattice momentum defined
in Eq.~(\ref{dismomt}) and $q$ is the
kinematical lattice momentum defined in Eq.~(\ref{latmomt}).
We can obtain $q$ numerically in this way from the tree-level massless
quark propagator.  We can compare this with the analytic form
for $q$ derived in the Appendix and given in Eq.~(\ref{q_mu}).

Note that for our mean-field improved Wilson-Dirac
operator, the tree-level limit for defining $q$ implies that we should
take $U\to I$ and $u_0\to 1$ in $D_{w}$ while keeping $\kappa$ fixed.
Thus the $\kappa$ that appears in the tree-level expression for
$q_\mu$ in Eq.~(\ref{q_mu}) is actually the improved $\kappa$
and not $\kappa^{\rm st}$.  This means that the tree-level
Wilson mass parameter, $m_w^{(0)}$, used in the Appendix is given by
$\kappa=1/[2(-m_w^{(0)}a)+(1/\kappa_c)]$ and hence differs from the $m_w$ in
Eq.~(\ref{kappa_defn}) used in the main body of the paper.   We have found
that the $q$ obtained in this way gives a much superior large momentum
behavior for the $M(p)$ and $Z(p)$ functions than is obtained 
when we do not use mean-field improvement.

Having identified the massless quark propagator in
Eq.~(\ref{massless_quark}), we can construct the massive overlap
quark propagator by simply adding a bare mass to its inverse,i.e.,
\begin{equation}
(S^{\rm bare})^{-1}(m^0)\equiv (S^{\rm bare})^{-1}(0)+m^0 \, .
\label{massive_inverse_prop}
\end{equation}
Hence, at tree-level we have for the massive, momentum-space
overlap quark propagator
\begin{equation}
S^{\rm bare}(m^0,p)\to S^{(0)}(m^0,p)=\frac{1}{iq\slh+m^0} 
\label{overlap_tree_prop}
\end{equation}
and the reason that the overlap quark propagator needs no tree-level
correction beyond identifying $q$ is now clear.

In order to complete the discussion we now relate our presentation
to standard notation used elsewhere.  We first define the dimensionless
overlap mass parameter
\begin{equation}
\mu \equiv \frac{m^0}{2m_w} 
\label{mu_defn}
\end{equation}
and then define $\tilde{D}_c^{-1}(\mu)$ in analogy with
Eq.~(\ref{massless_quark})
\begin{equation}
S^{\rm bare}(m^0)\equiv \tilde{D}_c^{-1}(\mu) \, ,
\label{overlap_propagator}
\end{equation}
i.e., $\tilde{D}_c^{-1}(\mu)$ is a generalization of $\tilde{D}_c^{-1}(0)$ 
to the case of nonzero mass.  Extending the analogy with the massless
case we introduce $\tilde{D}(\mu)$ and $D(\mu)$, which are generalized
versions of $\tilde{D}(0)$ and $D(0)$, through the definitions
\begin{equation}
\tilde{D}_c^{-1}(\mu) \equiv \frac{1}{2m_w} \tilde{D}^{-1}(\mu)
\hspace{1cm}{\rm{and}}\hspace{1cm}
\tilde{D}^{-1}(\mu) \equiv \frac{1}{1-\mu}\left[{D}^{-1}(\mu)-1\right]
\, .
\label{D_mu}
\end{equation}
We can now use Eq.~(\ref{massive_inverse_prop}) and the above definitions
to obtain an expression for $D(\mu)$.
From Eqs.~(\ref{massive_inverse_prop}), (\ref{mu_defn}) and
(\ref{D_mu}) we see that we must have
$\tilde D(\mu) = \tilde D(0) + \mu$.  Inverting this gives
\begin{equation}
\frac{1}{1-\mu}\left[D^{-1}(\mu)-1\right] =
           \left[ \tilde D(0)+\mu\right]^{-1}
\end{equation}
and so
\begin{equation}
D^{-1}(\mu) = (1-\mu)\left[  \tilde D(0)+\mu \right]^{-1} +1
            = \left[  \tilde D(0)+1 \right] 
                     \left[  \tilde D(0)+\mu \right]^{-1} \, .
\end{equation}
Inverting gives
\begin{equation}
D(\mu) = \left[  \tilde D(0)+\mu \right] \left[\tilde D(0)+1\right]^{-1}
\hspace{1cm}{\rm{and~also}}\hspace{1cm}
D(0) = \tilde D(0) \left[\tilde D(0)+1\right]^{-1}
\end{equation}
and so finally
\begin{eqnarray}
D(\mu) &=&\left[ (1-\mu)\tilde D(0)+
         \mu\left(\tilde D(0)+1\right)\right]
         \left[\tilde D(0)+1\right]^{-1}  \nonumber \\ 
       &=& (1-\mu)D(0)+\mu 
        = \frac{1}{2}\left[1+\mu+(1-\mu)\gamma_5 H_a\right]
    \, .
\label{D_mu_eqn}
\end{eqnarray}
We have then recovered the standard expression for $D(\mu)$ found for
example in Ref.~\cite{edwards2} and elsewhere.

We see that the massless limit, $m^0\to 0$, implies that $\mu\to 0$
and $D(\mu)\to D(0)$, $\tilde D^{-1}(\mu)\to \tilde D^{-1}(0)$ and
$\tilde{D}_c^{-1}(\mu)\to \tilde{D}_c^{-1}(0)$.  For non-negative
bare mass $m^0$ we require $\mu\geq 0$. In order that the above expressions
and manipulations be well-defined we must have $\mu<1$.  Hence,
$0\leq\mu<1$ defines the allowable range of bare masses.

Our numerical calculation begins with an evaluation of the
inverse of $D(\mu)$, where $D(\mu)$ is defined in
Eq.~(\ref{D_mu_eqn}) and using $H_a=\epsilon(H_w)$
for each gauge configuration in the ensemble.
We then calculate Eq.~(\ref{overlap_propagator}) for each
configuration and take the ensemble average to obtain
$S^{\rm bare}(x,y)$.  The discrete Fourier
transform of this finally gives the momentum-space bare quark propagator,
$S^{\rm bare}(p)$, for the bare quark mass $m^0$.

Our calculations used $\kappa=0.19$ and $u_0=0.88888$ and since at tree level
$\kappa_c=1/8$, we then have $m_w a= 1.661$.
Recall that the lattice spacing  is $a=0.125$~fm and so
we have $a^{-1}=1.58$~GeV and $m_w= 2.62$~GeV.
We calculated at ten quark masses specified by
$\mu=$ $0.024$, $0.028$, $0.032$, $0.040$, $0.048$, $0.060$, $0.080$,
$0.100$, $0.120$, and $0.140$.
This corresponds to bare masses in physical units of 
$m^0=2\mu m_w =$ $126$, $147$, $168$, $210$, $252$, $315$,
$420$, $524$, $629$, and $734$~MeV respectively.

\section{NUMERICAL RESULTS}
\label{numerical}

\begin{figure}[t]
\centering{\epsfig{angle=90,figure=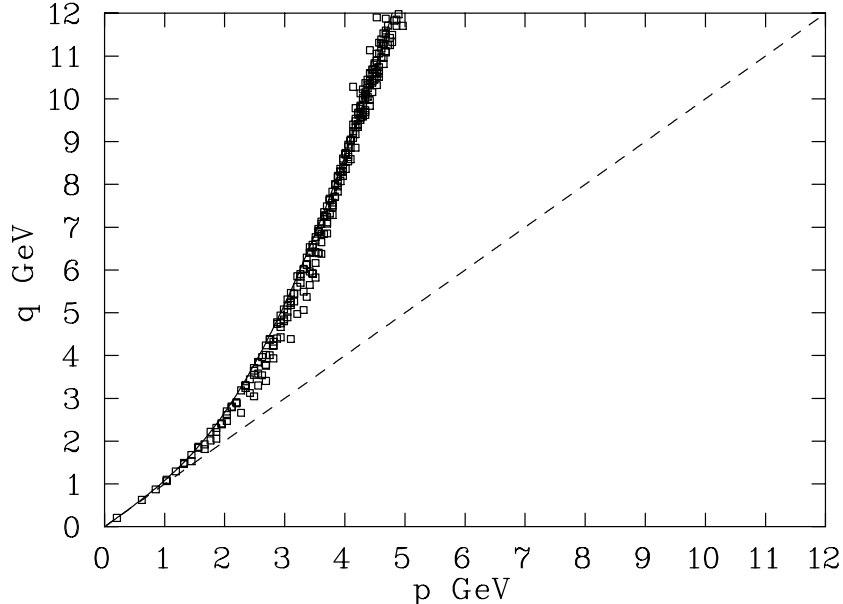,height=8cm} }
\parbox{130mm}{
    \caption{The kinematical momentum $q$ for overlap quarks versus the
     discrete momentum $p$ with both in GeV.  No data cuts have been applied.
     The analytic result from the Appendix for the case
     of purely diagonal momenta is shown as the solid line for
     comparison.         }
\label{pvsqfull} }
\end{figure}

We have numerically extracted the kinematical lattice
momentum $q_\mu$ directly from
the tree-level overlap propagator using
Eqs.~(\ref{curlyCandB}), (\ref{cmup_bp}) and (\ref{latmomt}).
In particular, by setting $U\to I$ and $u_0\to 1$
we have numerically verified to high-precision the
tree-level behavior in Eq.~(\ref{overlap_tree_prop}) for all ten of our
bare masses $m^0$, which is a good test of our code for extracting
the momentum-space quark propagator.
We plot $q\equiv \sqrt{\sum_\mu q_\mu^2}$ against
the discrete lattice momentum $p\equiv \sqrt{\sum_\mu p_\mu^2}$ in
Fig.~\ref{pvsqfull}.  For pure Wilson quarks we would have
$q_\mu= (1/a)\sin(p_\mu a)$.  It is interesting that $q$ for the overlap
lies above the discrete lattice $p$, while $q$ for Wilson quarks
lies below.  Of course in both cases, $q\to p$ at small $p$.

\subsection{Data Cuts and Averaging}

To clean up the data and improve our ability to draw conclusions about
continuum physics, we will on occasion employ the so-called
``cylinder cut'', where we select only lattice four-momentum lying
near the four-dimensional diagonal in order to minimize hypercubic
lattice artifacts.  This cut has been successfully used elsewhere
in combination with tree-level correction in studies of the quark and
gluon propagator~\cite{gluon_refs,jon1,jon2}.
It is motivated by the observation that for a given momentum squared,
($p^2$), choosing the smallest momentum values of each of the Cartesian
components, $p_\mu$, should minimize finite lattice spacing artifacts. 

\begin{figure}[t]
\centering{\epsfig{angle=90,figure=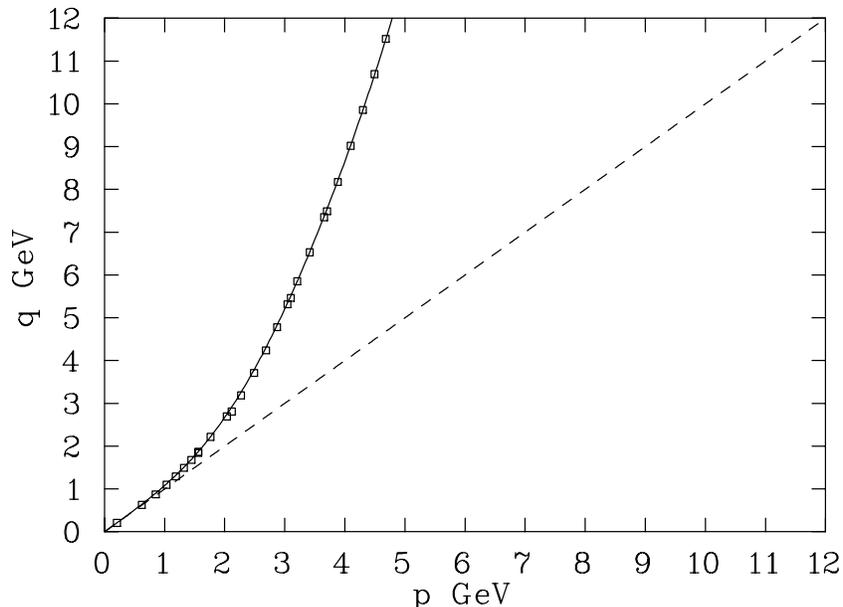,height=8cm} }
\parbox{130mm}{
    \caption{The kinematical momentum $q$ for overlap quarks versus the
     discrete momentum $p$ with both in GeV.  The cylinder
     cut has been applied and the hypercubic spread has been much
     reduced.  The analytic result from the Appendix for the case
     of purely diagonal momenta is shown as the solid line for
     comparison. 
             }
\label{pvsqcyl} }
\end{figure}

We calculate the distance 
$\Delta{p}$ of a momentum four-vector $p_\mu$ from the diagonal using
\begin{equation}
\Delta{p} = |p|\sin\theta_{p},
\end{equation}
where the angle $\theta_{p}$ is given by
\begin{equation}
\cos\theta_{p} = \frac{p \cdot \hat{n}}{|p|},
\end{equation}
and $\hat{n} = \frac{1}{2} (1,1,1,1)$ is the unit vector along the
four-diagonal.  For the cylinder cut employed in this study we neglect
points more than one spatial momentum unit $2\pi/N_i$
from the diagonal.  To see that this cut has the desired effect
of reducing hypercubic artifacts we plot the cut version of
Fig.~\ref{pvsqfull} in Fig.~\ref{pvsqcyl}.  The cylinder-cut data
points have much reduced hypercubic spread and lie on a
smooth curve.  We also sometimes make use of a ``half-cut'' where
we only retain momentum components $p_\mu$ half way out into the
Brillouin zone.

On an isotropic four dimensional lattice we have $Z(4)$ invariance. Since
our lattice is twice as long in the time-direction as it is in the spatial
direction, this symmetry is broken down to $Z(3)$.
This symmetry may be used to improve the statistics by averaging
over $Z(3)$-identical momentum points.  Since QCD and our lattice
are parity invariant, we can also perform a reflection average
at the same time.  This average treats the negative
momentum combinations in the same way as the positive ones. 
In an obvious notation, if we wish to calculate some
quantity $S(1,2,3,4)$, then we calculate all of the quantities
$S(\pm i,\pm j,\pm k,\pm 4)$ for $i,j,k$ being any permutation
of $1,2,3$ and perform an average over all of these quantities.

We could also, in principle, average over all lattice starting points
in the calculation of the propagator, since we should have
translational invariance, i.e., 
$S(y,x)$ should be the
same for all equal $(y-x)$. However, this averaging is too
expensive to implement and we use a single starting point
$x_\mu$ and calculate $S(y,x)$ for all finishing points
$y_\mu$.
We obtain the Fourier transform $S(p)$ in the usual way with
\begin{equation}
S(p) \equiv \sum_x e^{-ip\cdot (y-x)}S(y,x) \, . 
\label{Fourier_trans}
\end{equation}

\subsection{Overlap Quark Propagator}
\label{overlap_results}

%
\begin{figure}[tp]
\centering{\epsfig{angle=90,figure=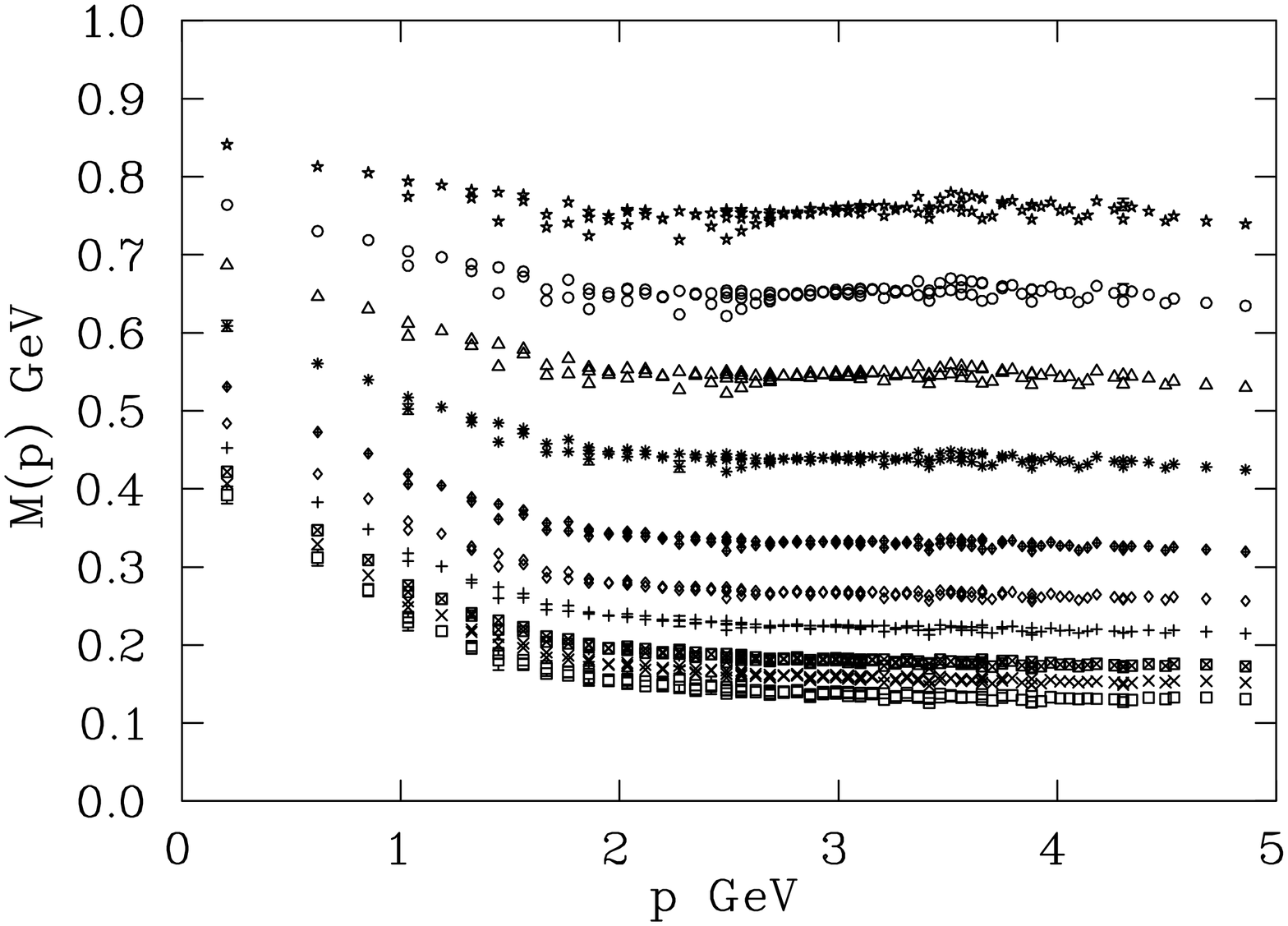,height=8cm} }
\centering{\epsfig{angle=90,figure=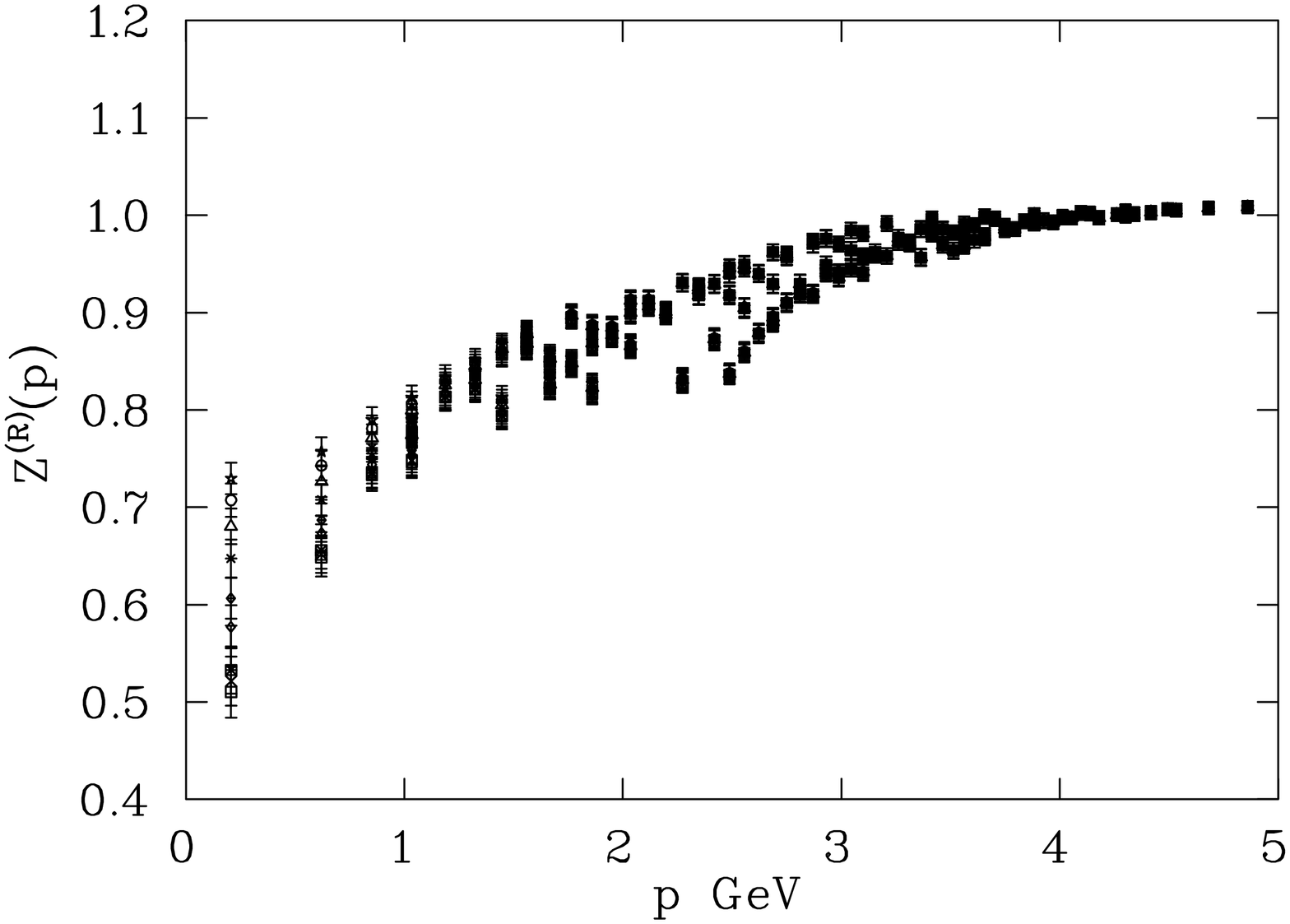,height=8cm} }
\parbox{130mm}{
  \caption{The functions $M(p)$ and 
$Z^{(\rm{R})}(p)\equiv Z(\zeta;p)$ for renormalization point
$\zeta=3.9$~GeV (on the $p$-scale) for all ten bare
quark masses for the half-cut data.  Data are plotted versus
the discrete momentum values defined in Eq.~(\ref{dismomt}), 
$p=\sqrt{\sum{p_\mu^{2}}}$, over the interval [0,5] GeV. 
The data in both parts of the figure correspond from
bottom to top to increasing bare quark masses, i.e.,
$\mu=$ $0.024$, $0.028$, $0.032$, $0.040$, $0.048$, $0.060$, $0.080$,
$0.100$, $0.120$, and $0.140$, which in physical units corresponds
to $m^0=2\mu m_w =$ $126$, $147$, $168$, $210$, $252$, $315$,
$420$, $524$, $629$, and $734$~MeV respectively.  The mass functions
at large momenta are very similar to the bare quark masses as
expected. 
}
\label{combmovrp}}
\end{figure}

In Fig.~\ref{combmovrp} we first show the half-cut results for all ten
masses for both the mass and wavefunction renormalization functions, 
$M(p)$ and $Z^{(\rm{R})}(p)\equiv Z(\zeta;p)$ respectively, against
the discrete lattice momentum $p$.  Statistical uncertainties are
estimated via a second-order, single-elimination
jackknife.
The renormalization point in Fig.~\ref{combmovrp} for $Z^{(\rm R)}(p)$ has been chosen as $\zeta=3.9$~GeV on the
$p$-scale.  We see that both $M(p)$ and $Z^{(\rm R)}(p)$ are reasonably
well-behaved up to 5~GeV although some anisotropy is evident.
We see that at large momenta the quark masses are approaching their
bare mass values as anticipated due to asymptotic freedom.

\begin{figure}[tp]
\centering{\epsfig{angle=90,figure=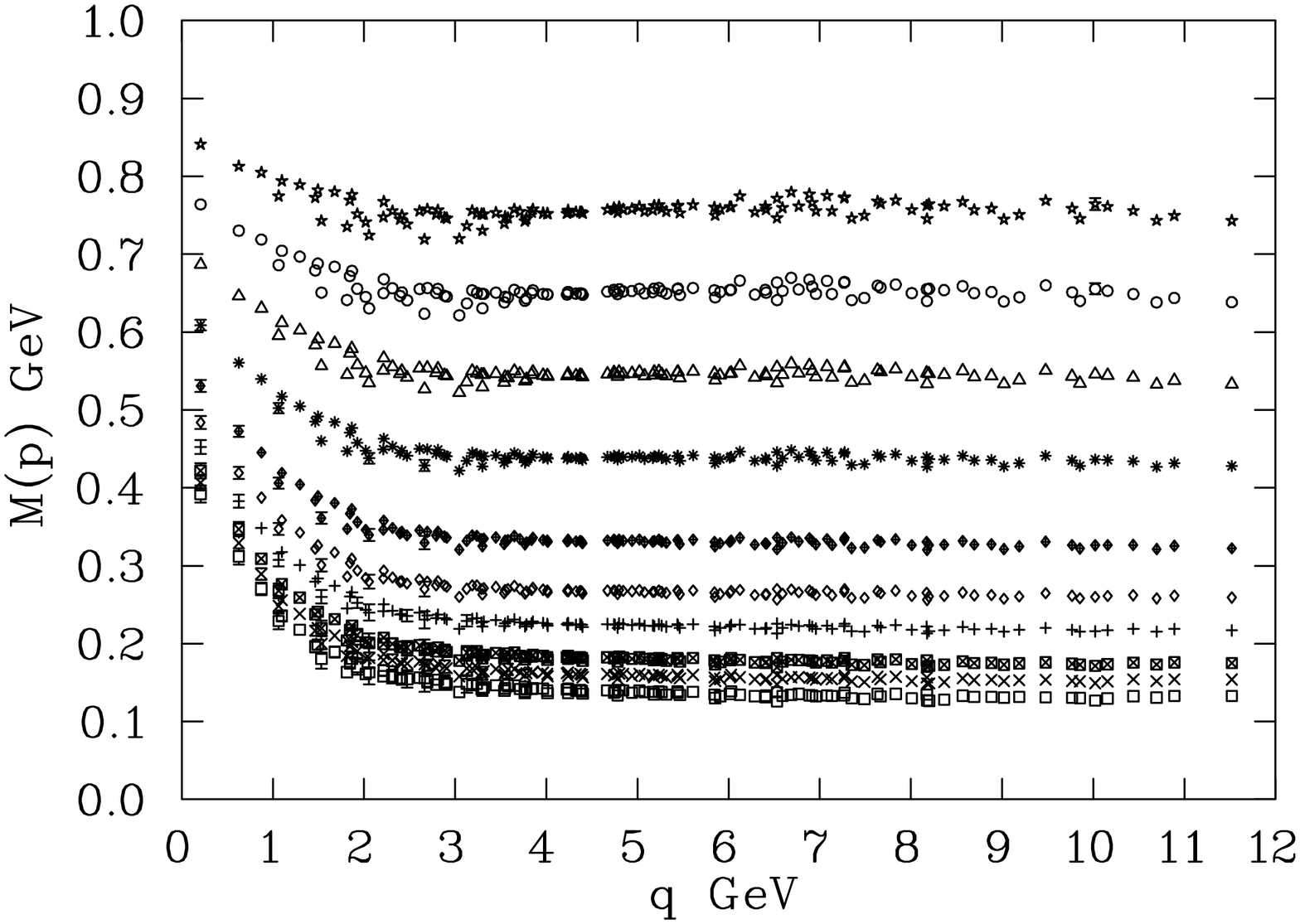,height=8cm} }
\centering{\epsfig{angle=90,figure=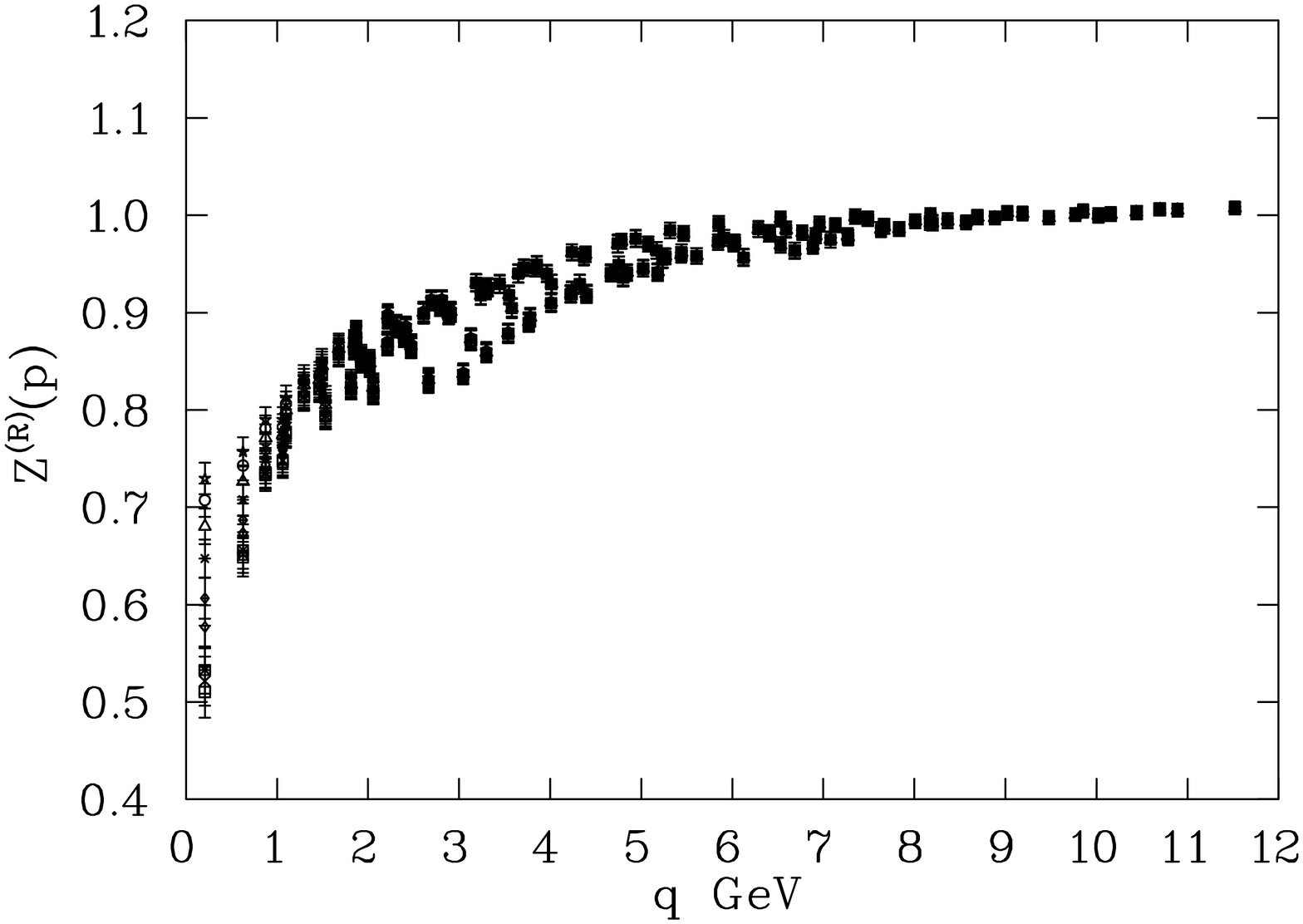,height=8cm} }
\parbox{130mm}{
  \caption{The functions $M(p)$ and 
$Z^{(\rm{R})}(p)\equiv Z(\zeta;p)$ for renormalization point
$\zeta=8.2$~GeV (on the $q$-scale) for all ten bare
quark masses for the half-cut data.  Data are plotted versus
the discrete momentum values defined in Eq.~(\ref{latmomt}), 
$q=\sqrt{\sum{q_\mu^{2}}}$, over the interval [0,12] GeV. 
The data in both parts of the figure correspond from
bottom to top to increasing bare quark masses.  The values of the bare
quark masses are in the caption of Fig.~\protect{\ref{combmovrp}}.
}
\label{combmovrq}}
\end{figure}

In the plots of $M(p)$ the data is ordered as one
would expect by the values for $\mu$, i.e., the larger the bare mass
$m^0$ the higher is the $M(p)$ curve.
In the figure for $Z^{(\rm R)}(p)$ the smaller the bare mass, the more pronounced
is the dip at low momenta.  Also, at small bare masses $M(p)$ falls off
more rapidly with momentum, which is understood from the fact that a
larger proportion of the infrared mass is due to dynamical chiral
symmetry breaking at small bare quark masses.
This qualitative behavior is consistent with what is seen
in Dyson-Schwinger based QCD models\cite{DSE_Review}.
The spread in the lattice data points indicates that some anisotropy from
hypercubic lattice artifacts has survived the identification of the
kinematical lattice momentum $q$.  In Fig.~\ref{combmovrq} we repeat
these plots but now against the kinematical lattice momentum $q$.
We see that the spread in the data is not significantly reduced and
that the kinematical momentum reaches up to 12~GeV.

\begin{figure}[tp]
\centering{\epsfig{angle=90,figure=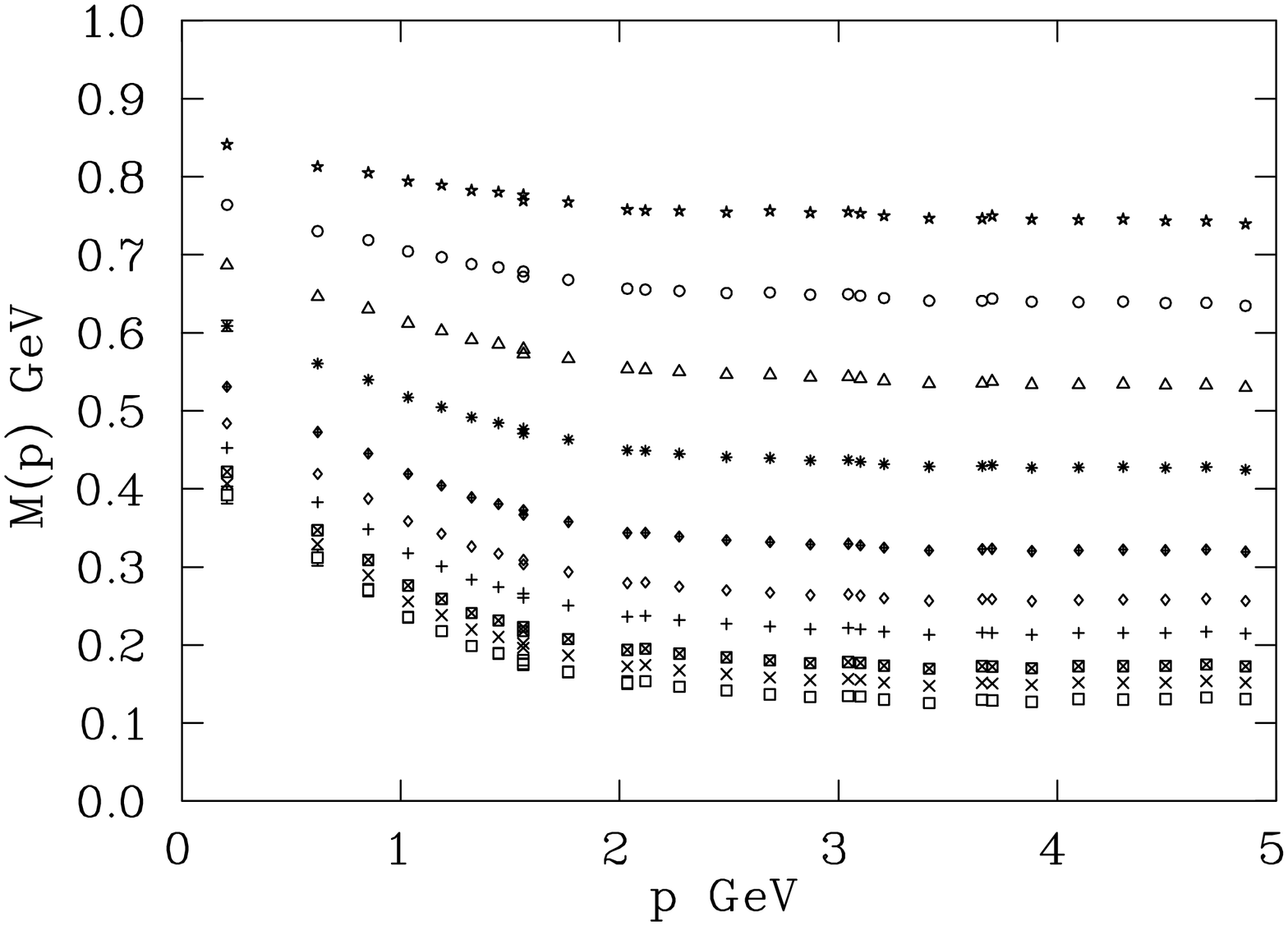,height=8cm} }
\centering{\epsfig{angle=90,figure=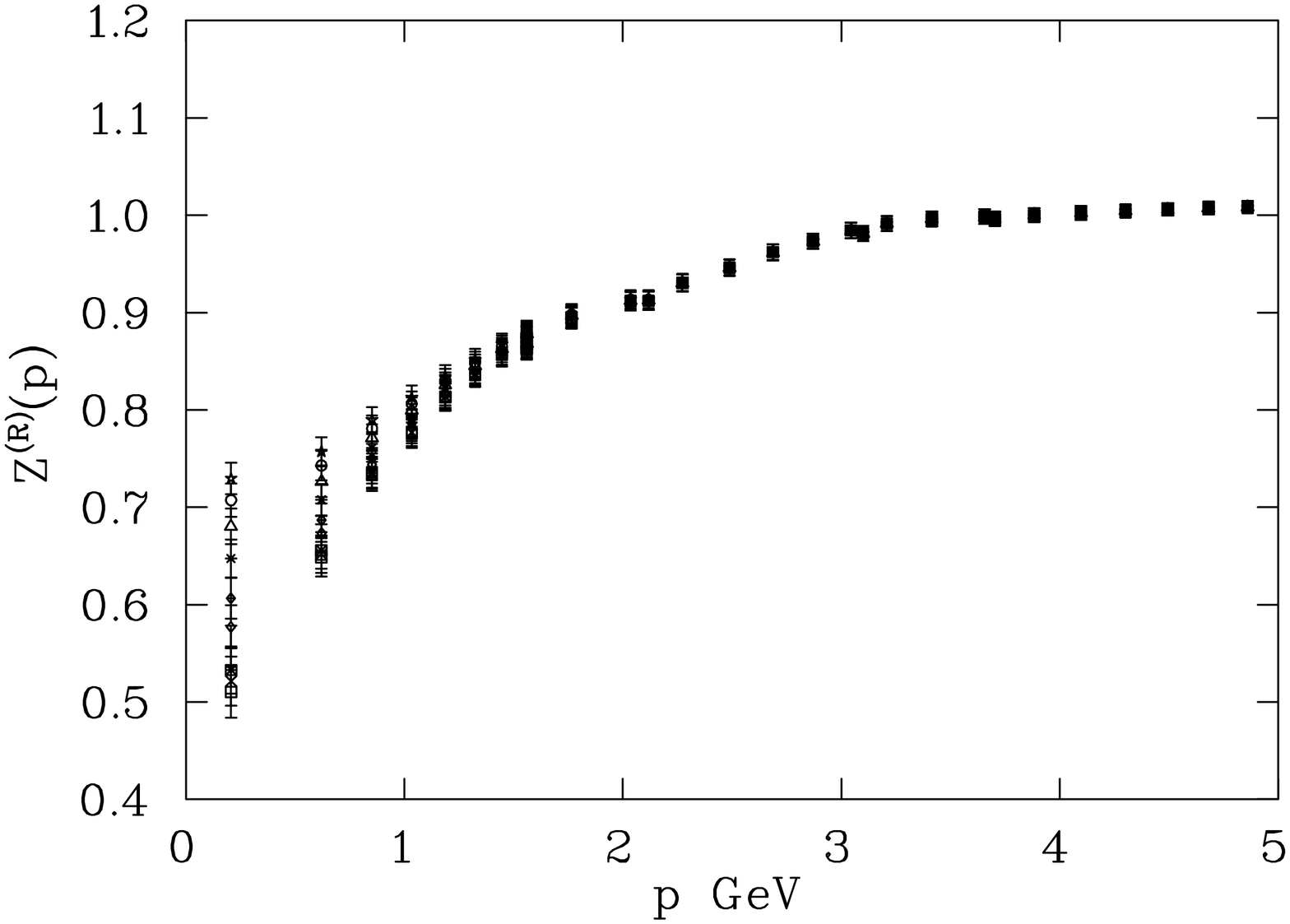,height=8cm} }
\parbox{130mm}{
  \caption{The functions $M(p)$ and 
$Z^{(\rm{R})}(p)\equiv Z(\zeta;p)$ for renormalization point
$\zeta=3.9$~GeV (on the $p$-scale) for all ten bare
quark masses and for data with a cylinder-cut, i.e.,
the data is identical to that of Fig.~\protect{\ref{combmovrp}}
except that it has been cylinder cut (one spatial momentum unit)
rather than half-cut.
}
\label{combmovr2pip}}
\end{figure}

\begin{figure}[tp]
\centering{\epsfig{angle=90,figure=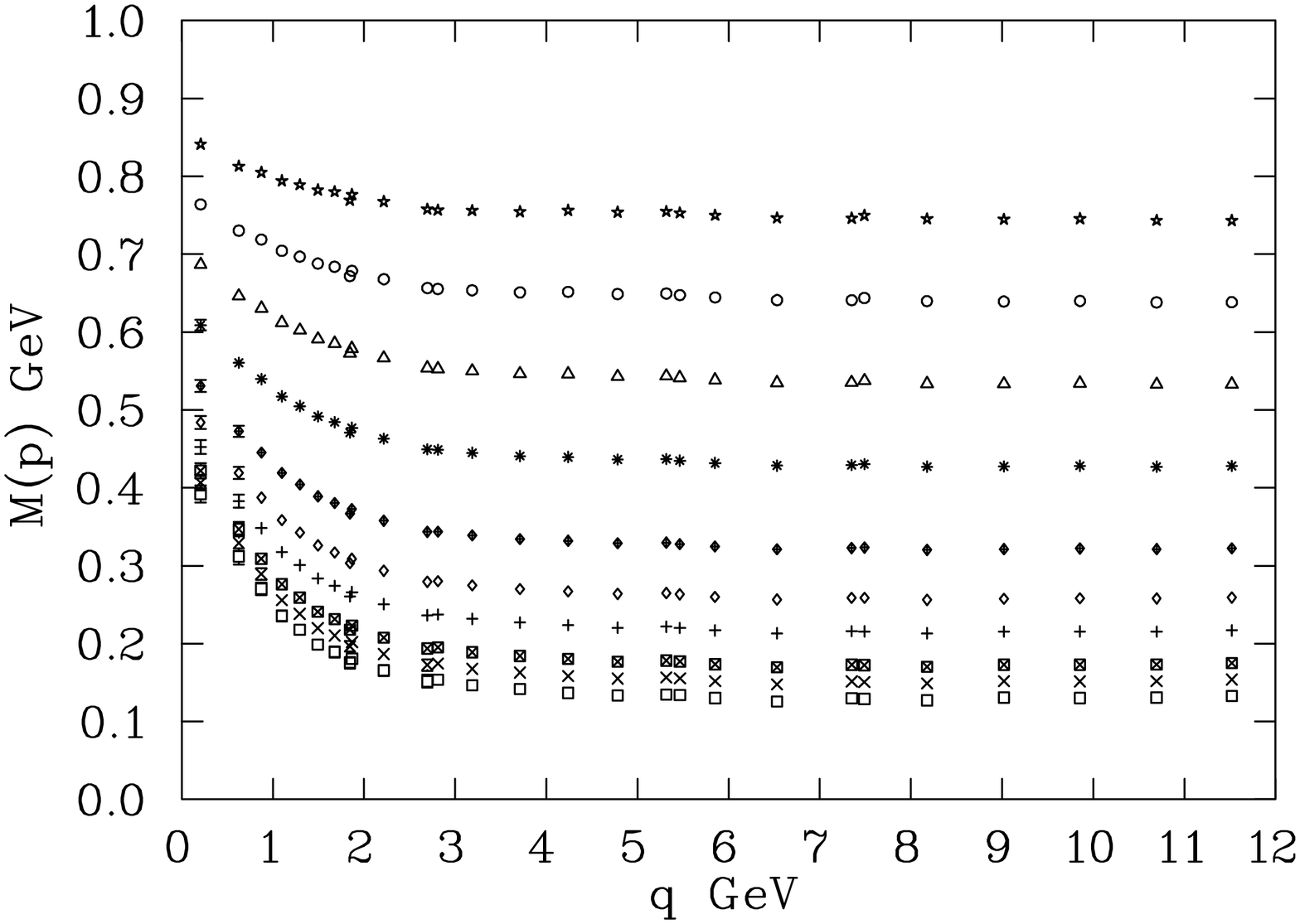,height=8cm} }
\centering{\epsfig{angle=90,figure=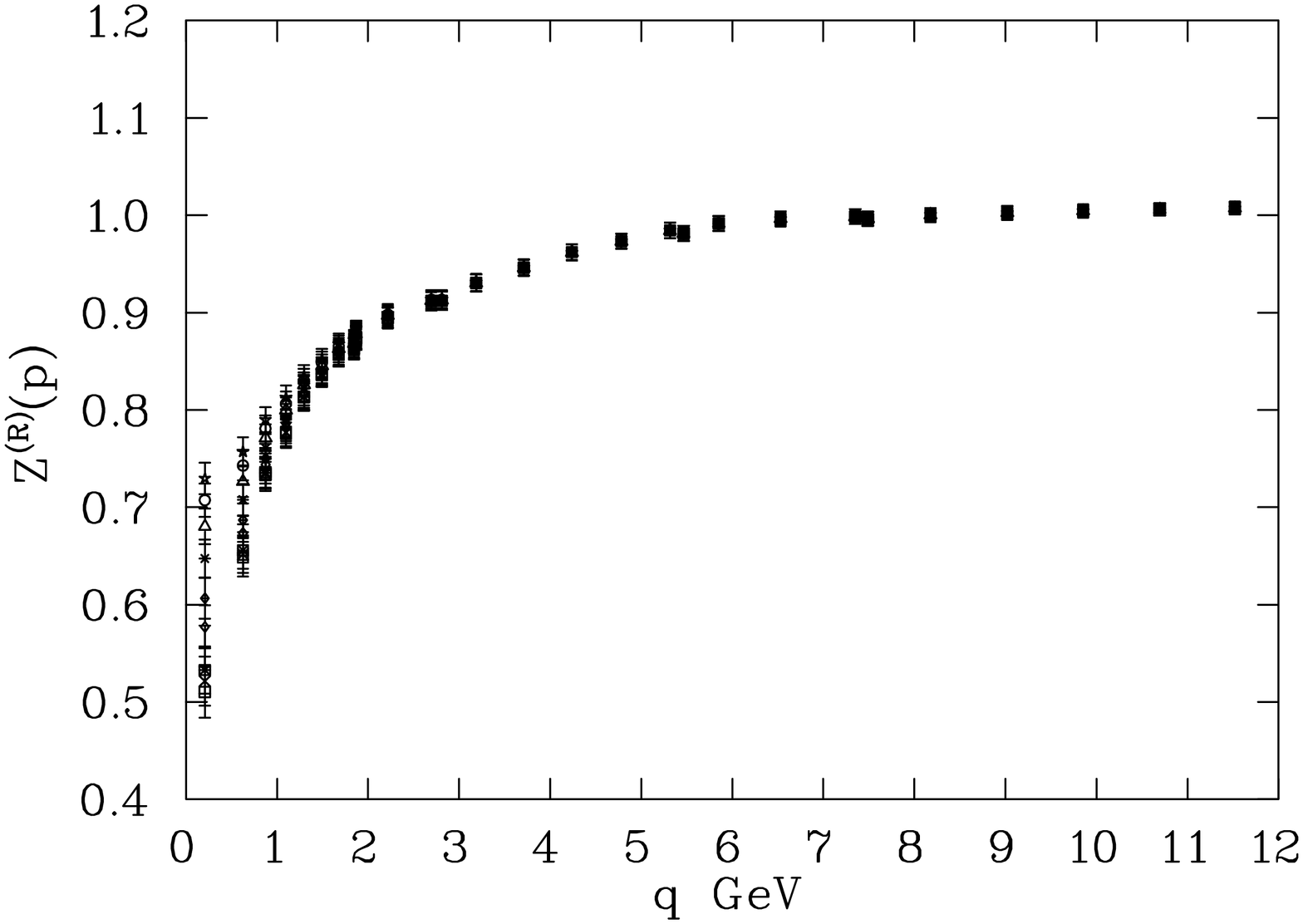,height=8cm} }
\parbox{130mm}{
  \caption{The functions $M(p)$ and 
$Z^{(\rm{R})}(p)\equiv Z(\zeta;p)$ for renormalization point
$\zeta=8.2$~GeV (on the $q$-scale) for all ten bare
quark masses and for data with a cylinder-cut, i.e.,
the data is identical to that of Fig.~\protect{\ref{combmovrq}}
except that it has been cylinder cut (one spatial momentum unit)
rather than half-cut.
}
\label{combmovr2piq}}
\end{figure}

The cylinder cut version of Figs.~\ref{combmovrp} and \ref{combmovrq} are given in
Figs.~\ref{combmovr2pip} and \ref{combmovr2piq} respectively.
The cylinder cut removes almost all of the remaining spread in the
overlap quark data and leaves data points which appear to lie on
smooth curves.  There is no apparent difference in the spread of
the cylinder-cut data when plotted against $p$ or $q$.
Experience with the gluon propagator\cite{gluon_refs}
suggests that the continuum limit for $Z(\xi,p)$ will be most
rapidly approached as $a\to 0$ by plotting it against its associated
kinematical lattice momentum $q$.  It is not obvious whether $M(p)$
would converge to its continuum-limit behavior more rapidly as $a\to 0$
by plotting against $q$ or $p$ or perhaps some other momentum scale.
The only way to resolve this question is to repeat the calculation
on a lattice with different spacing $a$ and to see which choice of
momentum on the horizontal axis leads to $a$-independent behavior
of $Z(\zeta;p)$ and $M(p)$ most rapidly as $a\to 0$. This is left for future
investigation.


\subsection{Extrapolation to Chiral Limit}

\begin{figure}[tp]
\centering{\epsfig{angle=90,figure=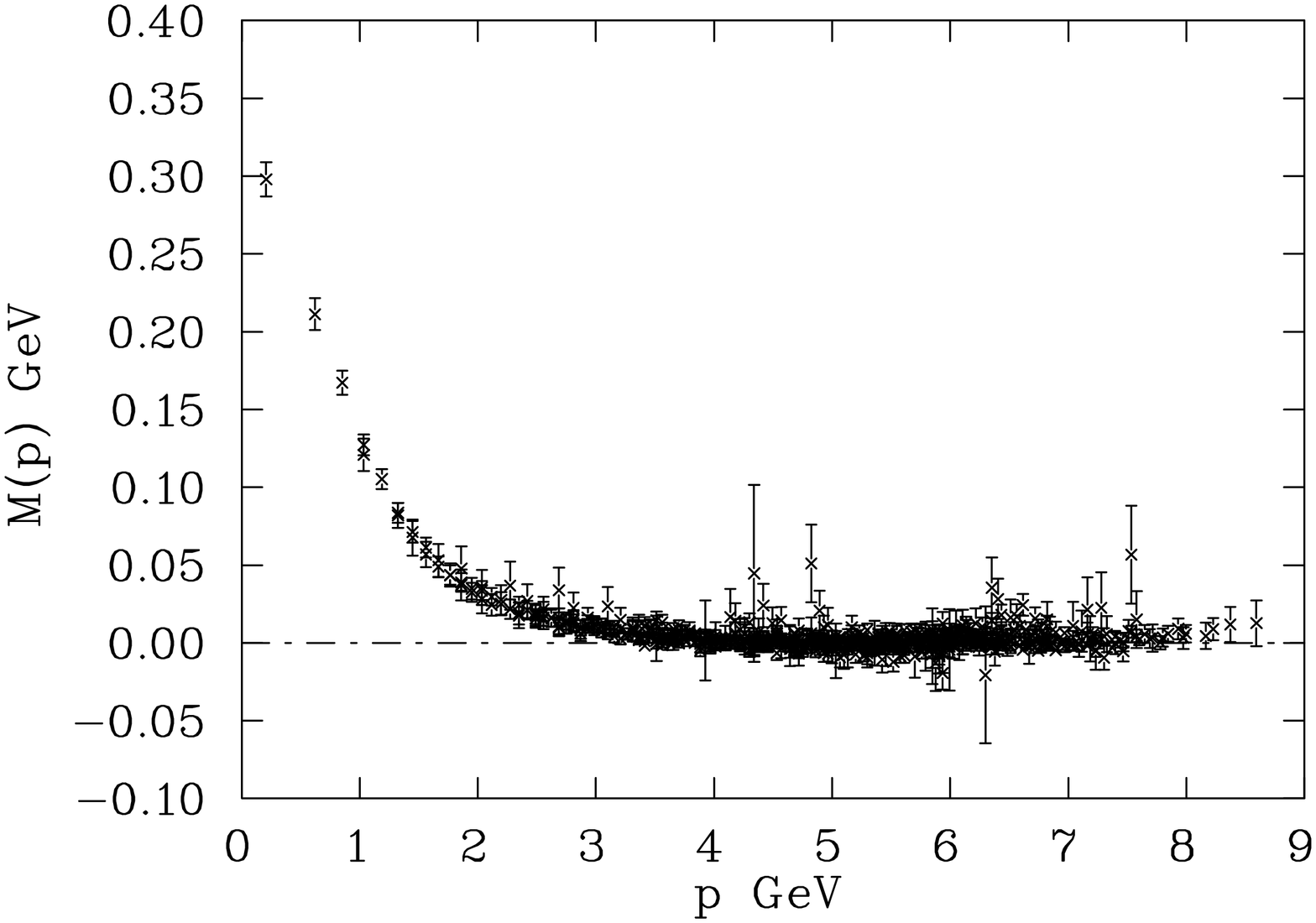,height=8cm} }
\centering{\epsfig{angle=90,figure=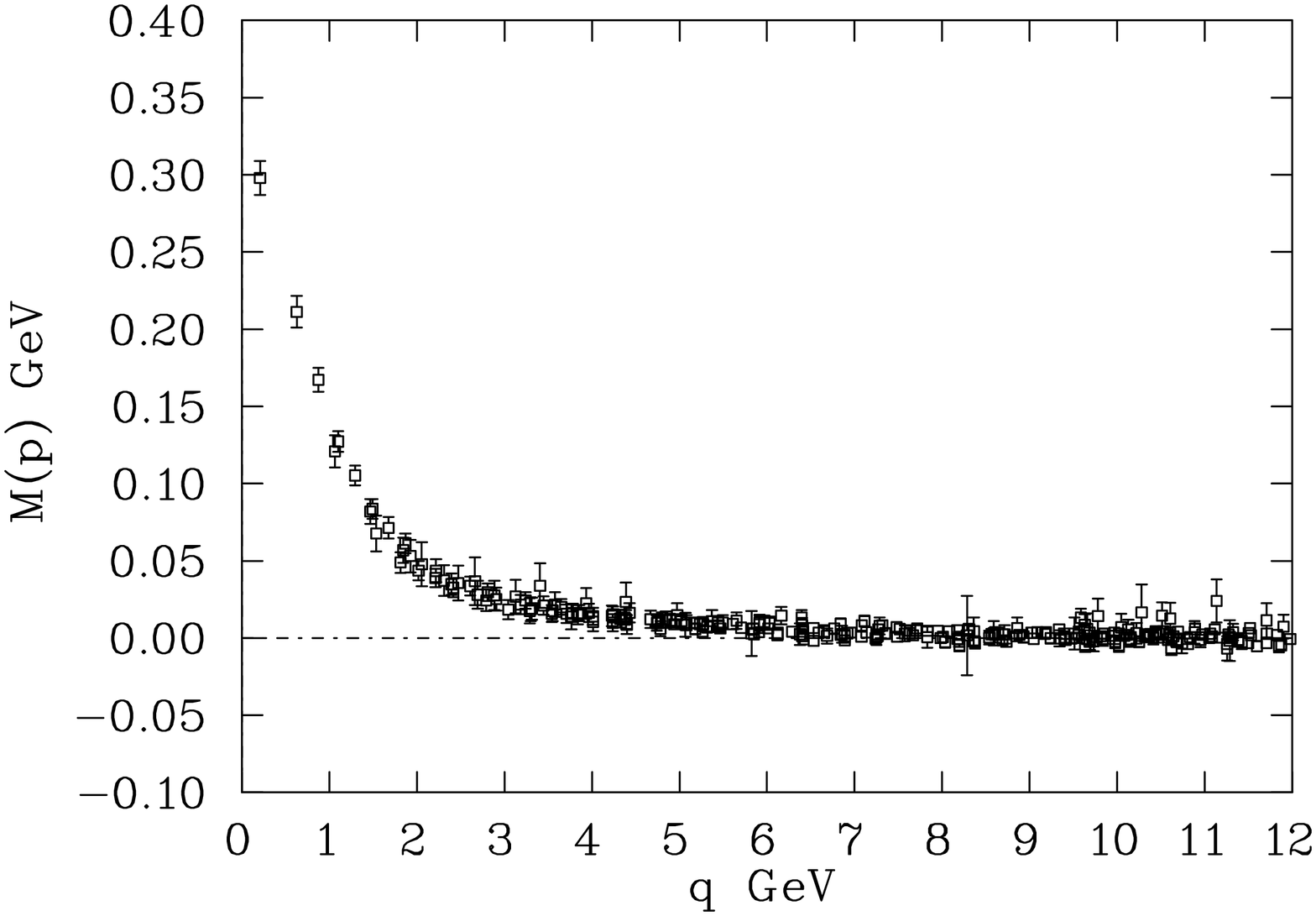,height=8cm} }
\parbox{130mm}{
\caption{The chiral limit mass function $M(p)$ obtained from
a simple linearly extrapolation of the various mass functions using the
full uncut data set.  This is plotted against both the discrete lattice
momentum $p$ and the kinematical lattice momentum $q$.  The latter is
shown only up to 12~GeV.}
\label{extram}}
\end{figure}
\begin{figure}[tp]
\centering{\epsfig{angle=90,figure=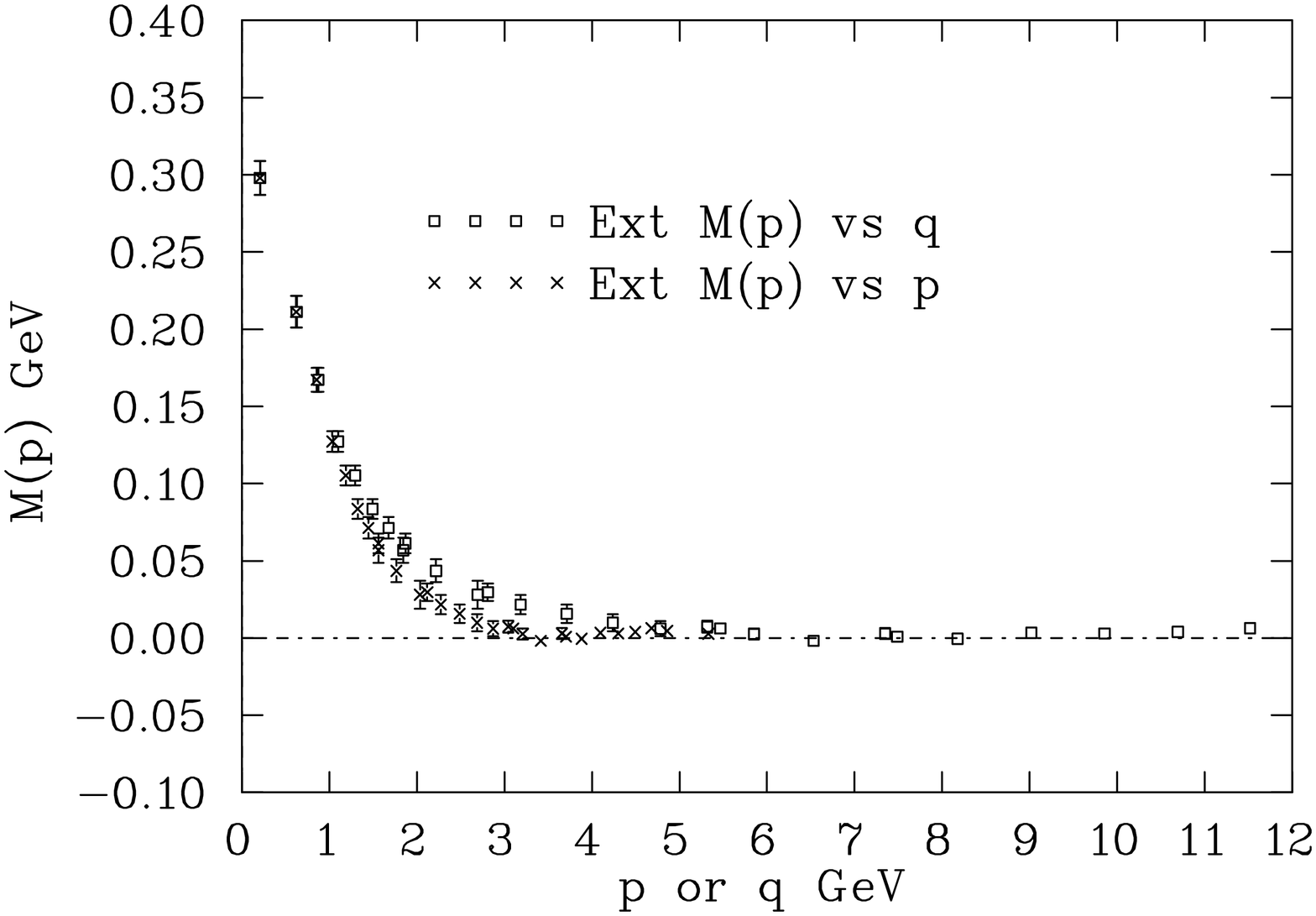,height=8cm} }
\centering{\epsfig{angle=90,figure=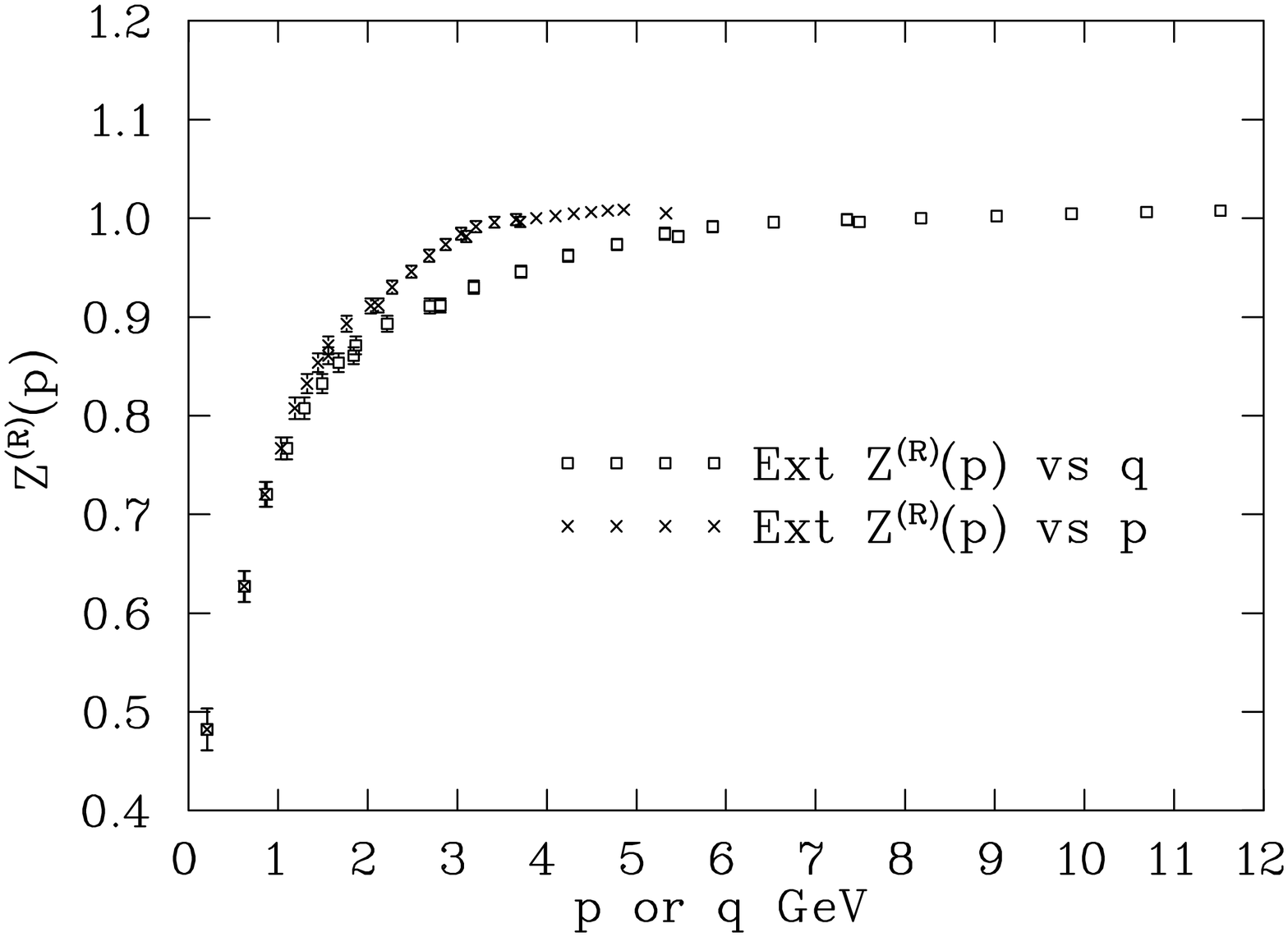,height=8cm} }
\parbox{130mm}{
\caption{The linearly extrapolated estimates of $M(p)$ and
$Z^{(\rm R)}(p)\equiv Z(\zeta;p)$ in the chiral limit
using the cylinder cut (one spatial momentum unit) data
of Figs.~\protect{\ref{combmovr2pip}}
and \protect{\ref{combmovr2piq}}. The values of the extrapolated
functions at the most infrared momentum point are
$M_{\rm IR}=297(11)$~MeV and $Z_{\rm IR}=0.48(2)$.}
\label{extramz2pi}}
\end{figure}

Our lightest bare quark mass is $m^0=126$~MeV and our heaviest
is 734~MeV and hence we expect that our results should
not be overly sensitive to the fact that our calculation is quenched.
For exploratory purposes, we regard our simulation results at our bare
quark masses to be a reasonable approximation to the infinite volume
and continuum limits.
We perform a simple linear extrapolation of our data to
a zero bare quark mass, i.e., a linear extrapolation to the chiral
limit. The results of our extrapolation for the mass function
are shown in  Fig.~\ref{extram}. The top figure shows the chiral
extrapolation of $M(p)$ for the full uncut data set and plotted against
$p$.  The bottom figure shows the same results plotted against $q$
up to 12~GeV.  The fact that the linear extrapolation gives an
$M(p)$ which vanishes within statistical errors at large momenta 
confirms that our simple linear extrapolation is reasonable at large
momenta. In fact, the data are found to be consistent with such a
linear fit at all momenta for the bare masses considered.

In Fig.~\ref{extramz2pi} we plot the cylinder-cut data
after the linear chiral extrapolation for both functions $M(p)$ and
$Z^{(\rm R)}(p)\equiv Z(\zeta;p)$.  These are shown against both
$p$ and $q$ with the renormalization points chosen as in the
previous figures, i.e., 3.9~GeV and 8.2~GeV for plots against
$p$ and $q$ respectively.  We see that both $M(p)$ and $Z^{(\rm R)}(p)$
deviate strongly from the tree-level behavior.  In particular,
as in earlier studies of the Landau gauge quark
propagator\cite{jon1,jon2,Bowman:2001xh}, we find 
a clear signal of dynamical mass generation and
a significant infrared suppression of the $Z(\zeta;p)$ function.
At the most infrared point, the dynamically generated mass has the value
$M_{\rm IR}=297(11)$~MeV and the momentum-dependent wavefunction
renormalization function has the value $Z_{\rm IR}=0.48(2)$.
The result $M_{\rm IR}=297(11)$~MeV is similar to typical mass
values attributed to the ``constituent quark'' mass and is
approximately 1/3 of the proton mass.
These values are very similar to the results found in previous
studies\cite{jon1,jon2,Bowman:2001xh} and are also similar to
typical values in QCD-inspired Dyson-Schwinger equation
models\cite{DSE_Review,DSE_stuff,Alkofer}. 

As the bare mass $m^0$ is increased from the chiral limit
as in Figs.~\ref{combmovr2pip} and \ref{combmovr2piq},
we are increasing the proportion
of explicit to dynamical chiral symmetry breaking.  Associated with
this we see that the dip
in $Z(\zeta;p)$ becomes less pronounced and the relative importance
of dynamical chiral symmetry breaking in the mass function $M(p)$
also decreases, i.e., we see that $M(p)$ becomes an increasingly
flat function of momentum as the bare mass is increased.


In the continuum at one loop order in perturbation theory and in the presence
of explicit chiral symmetry breaking (i.e., a nonzero bare mass),
the asymptotic behavior of the mass function is that of the running quark
mass.  Specifically for large, Euclidean $p^2$ and renormalization point
$\zeta$ we have the one--loop result~\cite{DSE_Review}
\begin{equation}
M(p^{2}) \, \stackrel{p^2,\zeta\to\infty}{\longrightarrow} \,
       m_\zeta \left[\frac{\ln(\zeta^{2}/\Lambda_{\rm QCD}^2)}
      {\ln(p^{2}/\Lambda_{\rm QCD}^2)}\right]^{d_M} \, ,
\label{runningmass}
\end{equation}
where $d_M=12/(33-2N_{f})$ is the anomalous mass dimension, $m_\zeta$
is the current quark mass, $N_f$ is the number of quark flavours, and
$\Lambda_{\rm QCD}$ is the QCD scale parameter..
In this limit we see then that the mass at the renormalization point,
$m(\zeta)\equiv M(\zeta^2)$, approaches the current quark mass, i.e.,
$m(\zeta)\equiv M(\zeta^2)\to m_\zeta$ as stated earlier.
The vanishing of the bare mass
$m^0$ defines the chiral limit and in that case the current quark mass
also vanishes and the asymptotic behavior of the mass function at one--loop
becomes
\begin{equation}
M(p^{2}) \, \stackrel{p^2,\zeta\to\infty}{\longrightarrow} \,
         \frac{4\pi^2 d_M}{3}
         \frac{\left(-\langle \bar qq\rangle_\zeta\right)}
                   {\left[\ln(\zeta^2/\Lambda_{\rm QCD}^2)\right]^{d_M}}
         \frac{1}{p^2}\left[\ln(p^2/\Lambda_{\rm QCD}^2) \right]^{d_M-1}
         \, .
\label{massfit}
\end{equation}
This is the asymptotic behavior of the dynamically generated quark mass.
We see that the running mass in Eq.~(\ref{runningmass}) falls off
logarithmically with momentum, whereas from 
Eq.~(\ref{massfit}) the dynamically generated mass falls off more rapidly
(as $1/p^2$ up to logarithms) in the chiral limit.
This is the reason that the effects of dynamical chiral symmetry breaking
can be neglected at large momenta. 
Since $M(p^2)$ is renormalization point independent, the combinations
$m_\zeta [\ln(\zeta^{2}/\Lambda_{\rm QCD}^2)]^{d_M}$,
$\langle \bar qq\rangle_\zeta/
\left[\ln(\zeta^2/\Lambda_{\rm QCD}^2)\right]^{d_M}$, and
$m_\zeta \langle \bar qq\rangle_\zeta$ are renormalization
group invariant.
The anomalous dimension of the quark propagator itself vanishes
in Landau gauge.   Hence in the continuum in Landau gauge
\begin{equation}
Z(\zeta;p^2) \stackrel{p^2,\zeta\to\infty}{\longrightarrow} \, 1 \, .
\label{asymptotic_Z}
\end{equation}

In our lattice results we clearly observe that
$Z^{(\rm R)}(p)\equiv Z(\zeta;p)$ behaves in a way consistent
with Eq.~(\ref{asymptotic_Z}).
We can then also examine whether or not the asymptotic behavior of our
linearly extrapolated chiral mass function satisfies Eq.~(\ref{massfit}).
Since we are working in the quenched approximation we have $N_f=0$.
We attempt
to extrapolate the quark condensate for three different values
of $\Lambda_{\rm QCD}$, i.e., 200, 234, and 300~MeV. Note that 234 MeV is among
typical values quoted for quenched QCD~\cite{ringwald}.

We also do the extraction over different ultraviolet fitting windows in order to
verify the insensitivity of the
chiral condensate to the fitting window. 
Since it is at present unclear whether $M(p)$ most rapidly approaches
the continuum limit when plotted against $p$ or plotted against $q$,
we have performed the fit to both, i.e., we have fitted
Eq.~(\ref{massfit}) to both sets of ultraviolet data for $M(p)$
in the half-cut version of the data in Fig.~\ref{extram}.

A summary of the fitting results is shown in Tables~\ref{qqbartabp} and
\ref{qqbartabq} for various fitting regions and QCD scale parameters.
As is standard practice, we quote the extracted condensate at the
renormalization scale $\zeta=1$~GeV using the renormalization
scale independence of  $\langle \bar qq\rangle_\zeta/
\left[\ln(\zeta^2/\Lambda_{\rm QCD}^2)\right]^{d_M}$.  It is the latter
that is extracted in the fit to the chiral $M(p)$. The extracted condensate is relatively insensitive to the value
of $\Lambda_{\rm QCD}$ and the fitting window.  There is however a very strong dependence on
which momentum scale is used for $M(p)$, i.e.,
$\sim 350$~MeV for $p$ compared with $\sim 600$~MeV for $q$.
It is clear that a quantitatively meaningful extraction of the quark
condensate will require us to establish which momentum
scale for $M(p)$ most rapidly reproduces the continuum limit as $a\to 0$.  
Other attempts\cite{Hernandez:1999cu,Hernandez:2001yn,DeGrand:2001ie} 
to directly calculate the quark condensate in the overlap
formalism in quenched QCD suggest a condensate value $\sim 250$~MeV,
which implies that $M(p)$ may be more appropriately plotted against
the discrete lattice momentum $p$.  This resolution of this issue
is beyond the scope of the present study and is
left for future work. However, once the correct momentum scale is identified
and the continuum limit estimated, the good quality of the overlap data
indicate that an extraction of the quark condensate should be possible.

\begin{table}
\caption{Summary of the results for the quark condensate, $-\left<\overline{q}q\right>_{\zeta}^{1/3}$, extracted from Eq.~(\ref{massfit}) in MeV and scaled to the renormalization point $\zeta=1.0$~GeV. The fit was done using Eq.~(\ref{massfit}) for each of $\Lambda_{\rm QCD}=200,234,300$~MeV on various momentum windows using $M(p)$ plotted against the discrete lattice momentum $p$.}
\begin{tabular}{cddd}
          & \multicolumn{3}{c}{$\Lambda_{\rm QCD}$}                                                   \\
$p$ GeV   & \multicolumn{1}{c}{200 MeV}   & \multicolumn{1}{c}{234 MeV}  &\multicolumn{1}{c}{300 MeV} \\
\hline  
3-5       & 356(35)                       & 347(34)                      & 333(32)                    \\
4-5       & 352(69)                       & 344(67)                      & 330(64)                    \\
\end{tabular}
\label{qqbartabp}
\end{table}

\begin{table}
\caption{Summary of the results for the quark condensate, $-\left<\overline{q}q\right>_{\zeta}^{1/3}$, extracted from Eq.~(\ref{massfit}) in MeV and scaled to the renormalization point $\zeta=1.0$~GeV. The fit was done using Eq.~(\ref{massfit}) for each of $\Lambda_{\rm QCD}=200,234,300$~MeV on various momentum windows using the linearly extrapolated half-cut data for $M(p)$ plotted against the kinematical lattice momentum $q$.}
\begin{tabular}{cddd}
          & \multicolumn{3}{c}{$\Lambda_{\rm QCD}$}                                                   \\
$p$ GeV   & \multicolumn{1}{c}{200 MeV}   & \multicolumn{1}{c}{234 MeV}  &\multicolumn{1}{c}{300 MeV} \\
\hline
3-5       & 604(68)                       & 591(67)                      & 566(64)                    \\
3-7       & 600(66)                       & 587(65)                      & 562(62)                    \\
3-9       & 594(63)                       & 581(62)                      & 557(59)                    \\
\\
4-5       & 613(81)                       & 599(79)                      & 574(76)                    \\
4-7       & 600(71)                       & 586(70)                      & 563(67)                    \\
4-9       & 589(67)                       & 576(65)                      & 553(63)                    \\
\\
5-7       & 590(66)                       & 577(65)                      & 556(32)                    \\
5-9       & 577(41)                       & 563(62)                      & 541(60)                    \\
\end{tabular}
\label{qqbartabq}
\end{table}

\section{Summary and Conclusions}
\label{conclusion}

To the best of our knowledge, this is the first detailed study
of the Landau
gauge momentum-space quark propagator in the overlap formalism.
By construction, the overlap quark propagator needs no
tree-level correction beyond the identification of the appropriate
kinematical lattice momentum $q$.  The quality of the data in the overlap
formalism is seen to be far superior to that from earlier
studies\cite{jon1,jon2}, which use an ${\cal O}(a)$-improved
Sheikholeslami-Wohlert (SW) quark action with a tree-level mean-field
improved clover coefficient $c_{\rm sw}$.  In these earlier studies
it was found that careful tree-level correction schemes are essential and the resulting
corrected data remain of inferior quality.
The quality of the data for the improved
staggered quark action, the so-called `Asqtad'' action with 
${\cal O}(a^4)$ and ${\cal O}(a^2g^4)$ errors, is also
seen to be superior to the ${\cal O}(a)$-improved quark action
and these results~\cite{Bowman:2001xh} are qualitatively consistent 
with what we have found here.

We use ten different quark masses and observe an approximately
linear relation  between the bare mass and the current quark mass for bare
masses in the range $\sim 125$ to $\sim 730$~MeV.  This allows a simple
linear extrapolation to the chiral limit.  Such a linear extrapolation
is justified in the ultraviolet, since the resulting
ultraviolet mass function vanishes within errors in the chiral limit
as expected.  For the most infrared
momentum points in the chiral limit using this linear extrapolation we find $M_{\rm IR}=297(11)$~MeV and
$Z_{\rm IR}=0.48(2)$ for the mass function and the momentum-dependent
wavefunction renormalization function respectively.

An extraction of the quark condensate from the ultraviolet
behavior of the chiral extrapolated mass function is possible with this
quality of data.  However, it is clear that this can not be done
in a quantitatively reliable way until one or more additional lattice
spacings become available so that we can identify the appropriate
momentum against which to plot $M(p^2)$.

The first calculation presented here is performed on a relatively small
volume of $1.5^3\times 3.0$~fm$^4$ and an intermediate lattice spacing
of 0.125~fm.  Ultimately, a variety of lattice spacings and volumes
should be used so that a study of the infinite-volume, continuum limit
of the overlap quark propagator can be performed.  It will also be
interesting to simulate at lighter quark masses in order to study the
chiral limit of the quenched theory in some detail.  Finally, one
should consider kernels in the overlap formalism other than
the pure Wilson kernel, e.g.,
using a fat-link irrelevant clover (FLIC) action\cite{Zanotti:2001yb}
as the overlap kernel\cite{Kamleh}.
These studies are currently underway and will be reported elsewhere.

\section{Acknowledgement}
Support for this research from the Australian Research Council
is gratefully acknowledged.  POB was in part supported by DOE contract
DE-FG02-97ER41022.  This work was carried out on the Orion supercomputer
and on the CM-5 at Adelaide University.  We thank Paul Coddington and
Francis Vaughan for 
supercomputer support.


\appendix

\section{Tree-level behavior}
\label{appendix}

\subsection{Tree-level overlap propagator}
\label{tre-ovelap}

We can derive an explicit form for the tree-level (i.e., free)
overlap quark propagator 
with the Wilson fermion kernel. Let us work in the infinite volume limit
with finite lattice spacing $a$.

It is convenient to define the dimensionless momentum variables
\begin{equation}
\tilde{k}_{\mu} \equiv \sin(p_{\mu}a ) \, ,\hskip1.5cm
\hat{k}_{\mu} \equiv 2\sin(p_{\mu}a/2) \, .
\end{equation}
Let us also define the dimensionless combination
\begin{equation}
A \equiv \left[ (-am_w^{(0)}) + \frac{r}{2} \hat{k}_{\mu}^2 \right] \, ,
\end{equation}
where $(-am_w^{(0)})$ is the negative, dimensionless tree-level
Wilson mass defined by $\kappa\equiv 1/[2(-m_w^{(0)}a)+(1/\kappa_c)]$.
We have $\kappa_c=1/8$ 
and $r$ is the usual Wilson parameter,  (typically one chooses $r=1$). 
Note that for small momenta we have $ A < 0 $ .
We can then write the momentum-space Wilson operator at tree-level as
\begin{equation}
D_w = \frac{1}{2\kappa}\left(i \gamma\cdot \tilde{k} + A \right) \, .
\end{equation}
It follows that
\begin{equation}
\sqrt{D^\dagger_w D_w} = \frac{1}{2\kappa} \sqrt{\tilde{k}^2 + A^2} \, ,
\label{Dw_root}
\end{equation}
where it is to be understood that by definition 
{\em only the positive root} is kept.  In Euclidean space it is
clear that we will always have $\tilde{k}^2 + A^2>0$ and the square root
is always well-defined.
The momentum-space overlap Dirac operator can then be written as
\begin{eqnarray}
D(0) & \equiv & \frac{1}{2}\left[1 + \gamma_5 H_a \right]   
= \frac{1}{2}\left[1 + \frac{D_w}{\sqrt{D^\dagger_w D_w}}\right]
 \nonumber \\
& = & \frac{1}{2} \left[
     1 + \frac{i\gamma\cdot\tilde{k} + A}
              {\sqrt{\tilde{k}^2 + A^2}}
                  \right]  \nonumber \\
& = & \frac{1}{2}\left[
         \frac{
      i\gamma\cdot\tilde{k} + \left\{A + \sqrt{\tilde{k}^2 + A^2}\right\}}
       { \sqrt{\tilde{k}^2 + A^2} }
                 \right] \, .
\label{D0_tree}
\end {eqnarray}
We see that $ H_a = \gamma_5 D_w/\sqrt{D^\dagger_w D_w}$ and that $ H^\dagger_a = H_a $  and 
$H_a^2 = 1$ as required in the overlap formalism, i.e., 
$H_a $ has eigenvalues $\pm$ 1 . 
We can readily invert $D(0)$ to give
\begin{eqnarray}
D^{-1}(0) & = & 2 {\sqrt{\tilde{k}^2 + A^2}}
   \left[\frac{-i\gamma\cdot\tilde{k}
   + \left\{  A + \sqrt{\tilde{k}^2 + A^2}\right\}}
   {\tilde{k}^2 +
   \left\{  A + \sqrt{\tilde{k}^2 + A^2}\right\}^2  }\right]  \nonumber \\
 & = &
   \left[ \frac{-i\gamma\cdot\tilde{k}}{ A + \sqrt{\tilde{k}^2 + A^2}} + 1 
   \right] 
\end {eqnarray}
and then
\begin{equation}
\tilde{D}^{-1}(0) \equiv D^{-1}(0) -1 = \frac{-i\gamma\cdot\tilde{k}}
{ A + \sqrt{\tilde{k}^2 + A^2}}
 = \frac{\tilde{k}^2}{i\gamma\cdot\tilde{k} 
\left\{  A + \sqrt{\tilde{k}^2 + A^2}\right\}} \, .
\end {equation} 
Clearly then $\{ \tilde{D}^{-1}(0), \gamma_5 \}  = 0$ as it must in the overlap formalism. The tree-level
momentum-space overlap quark propagator in the massless limit is then given by
\begin{equation}
S^{(0)}(0,p) = \frac{1}{2m_w^{(0)}}\tilde{D}^{-1}(0) 
 = \frac{1}{2m_w^{(0)}} \left[\frac{\tilde{k}^2}{i\gamma . \tilde{k} 
 \left\{  A + \sqrt{\tilde{k}^2 + A^2}\right\}}\right] 
 \equiv  \frac{1}{i{q\slh}} \, .
\end{equation}
Hence, we recognize that the kinematical tree-level momentum is given by
\begin{equation}
q_{\mu} = {2m_w^{(0)}} \tilde{k}_{\mu}\frac{ \left\{A + \sqrt{\tilde{k}^2 + A^2}\right\}}{\tilde{k}^2} \,.
\label{q_mu}
\end{equation}
This analytic form for the
$q\equiv \sqrt{\sum q_\mu^2}$ versus 
$p\equiv \sqrt{\sum p_\mu^2}$ behavior
is plotted as the solid line in Figs.~\ref{pvsqfull} and \ref{pvsqcyl}
for the case of purely diagonal momenta, ($p_1=p_2=p_3=p_4$).  The analytic
form can be checked against each (diagonal or non-diagonal) point on
a case-by-case basis and they agree to within numerical precision.

We can verify analytically that $ q_{\mu} \to p_{\mu}$ as $p \to 0$ as seen
in Fig.~\ref{pvsqcyl} .  Note that as
$p \to 0$ we have $ A < 0$ and hence
\begin{eqnarray}
q_{\mu} & \rightarrow & {2m_w^{(0)}} \tilde{k}_{\mu}\frac{ \sqrt{\tilde{k}^2 + |A|^2}  - |A|}{\tilde{k}^2 } \nonumber  \\
   & \rightarrow & \tilde{k}_{\mu} \frac{m_w^{(0)}}{|A|} \rightarrow \frac{1}{a}\sin(p_{\mu}a) \rightarrow p_{\mu} \,.
\label{lim_q_to_p}
\end {eqnarray}

\subsection{Tree-level dispersion relation}
 
The massless, tree-level overlap propagator has the momentum-space
form
\begin{equation}
S^{(0)}(0,p) = \frac{-i{q\slh}}{q^2}
\end{equation}
and so has poles when $ q^2 = 0$ .
We can analytically continue $ q_4 \to iE $ and then we find poles
at $E=|\vec{q}| $, 
i.e., in terms of our tree-level corrected propagator we have a perfect 
massless dispersion relation.

However, for hadronic properties without tree-level correction it is
the dispersion relation
in $p$ that is relevant, i.e., we need to analytically continue
$p_4 \to iE $ and find the poles
in $ S^{(0)}(0,p) $.  Our discussion here generalizes that given
in Ref.~\cite{Niedermayer:1998bi}.  It is clear from Eq.~(\ref{Dw_root}) and Eq.~(\ref{D0_tree})
that the analytic continuation is only defined in the region where
$\tilde{k}^2+A^2\geq 0$, since otherwise the argument of the square-root
is negative and the definition of $D(0)$ has no meaning.

The poles occur when
\begin{eqnarray}
0 & = & q^2 = 4(m_w^{(0)})^2 \frac{{\left\{  A + \sqrt{\tilde{k}^2 + A^2}
   \right\}}^2}{\tilde{k}^2}  \nonumber  \\
  & = & 4 (m_w^{(0)})^2 \left( 1 + \frac{ 2A^2}{\tilde{k}^2}
      \left[ 1 + {\rm sgn}(A) 
     \sqrt{ 1 + ({\tilde{k}^2}/A^2})\right]\right) \,.
\end{eqnarray}
Provided $ A < 0 $ we see that $ q^2 \to 0$ as $ {\tilde{k}^2} \to 0 $.
Consider these poles when 
$ p_4 \to iE $ and $ \vec{p}=(0,0,p)$, then the conditions $ {\tilde{k}^2} = 0 $ ,  $ A < 0 $
become
\begin{equation}
\sin^2(iEa) + \sin^2(pa) = 0 \, ,
\end{equation}
\begin{equation}
\sin^2(iEa/2) + \sin^2(pa/2) < \frac{am_w^{(0)}}{2r}  
\end{equation}
respectively. Thus we have poles at
\begin{equation}
\cosh(Ea) = \sqrt{ 1 + \sin^2(pa)}
\label{dispersion_eq}
\end{equation}
when we satisfy the condition
\begin{equation}
\cosh(Ea) > 2 - \cos(pa) - \frac{am_w^{(0)}}{r} \,.
\label{A_condition}
\end{equation}
Note that the analytic continuation to Minkowski space is only well-defined
when the square-root operation is well-defined, i.e., for
$\tilde{k}^2+A^2\geq 0$ and so this condition must also be satisfied.
We can rewrite this condition as
\begin{equation}
\cosh(Ea) \leq \frac{2+[2-(am_w^{(0)})]^2-2[2-(am_w^{(0)})]\cos(pa)}
                 {2[2-(am_w^{(0)})-\cos(pa)]}\,.
\label{k2A2_condition}
\end{equation}
Recall that Eq.~(\ref{A_condition}) is equivalent to the condition $A<0$.
If $ A = 0$ then 
$ q^2 = 4 (m_w^{(0)})^2 \ne 0 $ for
any real $\tilde{k}^2$ and hence there are no poles.
If $ A > 0$ then in the region where the square-root is well
defined $ \left\{  A + \sqrt{\tilde{k}^2 + A^2}\right\} > 0 $
and there are no poles in that case either.

\begin{figure}[t]
\centering{\epsfig{angle=90,figure=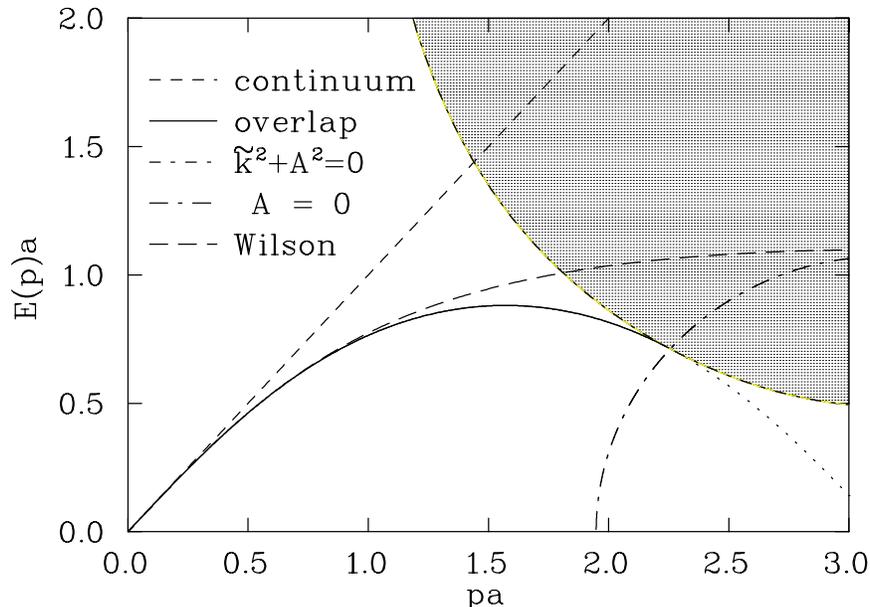,height=8cm} }
\parbox{130mm}{
    \caption{
The dispersion relation for the overlap quark propagator of
Eq.~(\protect\ref{dispersion_eq}) is shown as the solid line
and corresponds to $\tilde{k}^2=0$.
The dispersion relation does not continue into the region
where $A>0$, i.e., it does not extend to the right beyond
the long-dash dot line denoting $A=0$, [the solution of
Eq.~(\protect\ref{A_condition})].
The analytic continuation to Minkowski space has no meaning when
$\tilde{k}^2+A^2<0$, i.e., it is undefined
above the short-dash dot line, [i.e., the solution
of Eq.~(\protect\ref{k2A2_condition})].  The intersection point
of these three curves is where we simultaneously
have $A$, $\tilde{k}^2$, and $\tilde{k}^2+A^2$ equal to zero.
Also shown for reference are the
dispersion relations for the continuum limit (short dashes) and for the
ordinary Wilson action (long dashes).
             }
\label{dispersion} }
\end{figure}

\end{document}